\documentclass[aps,prd,preprint,nofootinbib,superscriptaddress]{revtex4}
\usepackage{color}
\usepackage[dvipdfmx]{graphicx}
\usepackage{psfrag}

\begin{document}

%%%%%%%%%%%%%%%%%%%%%% MACROS
\newcommand{\nn}{\nonumber}
\def\dfrac#1#2{\displaystyle\frac{#1}{#2}}
\newcommand{\ovl}[1]{\overline{#1}}
\newcommand{\wt}[1]{\widetilde{#1}}
\newcommand{\eq}[1]{Eq.~(\ref{#1})}
\newcommand{\eqn}[1]{(\ref{#1})}
\newcommand{\p}{\partial}
\newcommand{\pslash}{p\kern-1ex /}
\newcommand{\qslash}{q\kern-1ex /}
\newcommand{\lslash}{l\kern-1ex /}
\newcommand{\sslash}{s\kern-1ex /}
\newcommand{\kaslash}{k_a\kern-2ex /}
\newcommand{\kbslash}{k_b\kern-2ex /}
\newcommand{\Dslash}{\mathcal{D}\kern-1.5ex /}
\newcommand{\bpsi}{\overline{\psi}}
\newcommand{\bc}{\overline{c}}
\newcommand{\tr}{\mathrm{tr}}
\newcommand{\vev}[1]{\langle #1 \rangle}
\newcommand{\VEV}[1]{\left\langle\mathrm{T} #1\right\rangle}
\newcommand{\beqa}{\begin{eqnarray}}
\newcommand{\eeqa}{\end{eqnarray}}
\newcommand{\rmO}{\mathrm{O}}
\newcommand{\rmd}{\mathrm{d}}
\newcommand{\rme}{\mathrm{e}}
\newcommand{\ba}{\begin{eqnarray}}
\newcommand{\ea}{\end{eqnarray}}
\newcommand{\be}{\begin{equation}}   
\newcommand{\Nf}{N_\mathrm{f}}

%%%%%%%%%%%%%% END OF MACROS %%%%%%%%%%%%%%%%%%
\preprint{UTHEP-603}
\preprint{OU-HET-755}
%\draft
\title{Chiral symmetry restoration, eigenvalue density of Dirac operator and axial U(1) anomaly at finite temperature}
\newcommand{\TsukubaA}{
Graduate School of Pure and Applied Sciences, University of Tsukuba, Tsukuba, Ibaraki 305-8571, Japan  
}
\newcommand{\TsukubaB}{
Center for Computational Sciences, University of Tsukuba, Tsukuba, Ibaraki 305-8577, Japan  
}
\newcommand{\Osaka}{
  Department of Physics, Osaka University,
  Toyonaka 560-0043, Japan
}
\author{Sinya~Aoki}
\affiliation{\TsukubaA}
\affiliation{\TsukubaB}
\author{Hidenori~Fukaya}
\affiliation{\Osaka}
\author{Yusuke~Taniguchi}
\affiliation{\TsukubaA}
\affiliation{\TsukubaB}

\begin{abstract}
 We reconsider constraints on the eigenvalue density of the Dirac operator in the chiral symmetric phase of 2 flavor QCD at finite temperature.
To avoid possible ultra-violet(UV) divergences, we work on a lattice, employing the overlap Dirac operator, 
which ensures the exact ``chiral'' symmetry at finite lattice spacings.
Studying multi-point correlation functions in various channels 
and taking their thermodynamical limit  (and then taking the chiral limit), 
we obtain stronger constraints than those found in the previous studies:
both the eigenvalue density at the origin and its first and second derivatives vanish in the chiral limit of 2 flavor QCD.
In addition  we show that the axial U(1) anomaly becomes invisible in susceptibilities of scalar and pseudo scalar mesons, suggesting that the 2nd order chiral phase transition with the $O(4)$ scaling is not realized in 2 flavor QCD. 
Possible lattice artifacts when non-chiral lattice Dirac operator is employed are briefly discussed.
\end{abstract}

\keywords{ Chiral symmetry, U(1) symmetry, finite temperature, eigenvalue density}
\maketitle
%\dedicated{}

%\begin{document}

%%%%%%%%%%%%%%%%%%%%%%%%%%
%%%%%%%%%%%%%%%%%%%%%%%%%%
\section{Introduction}
%%%%%%%%%%%%%%%%%%%%%%%%%%
%%%%%%%%%%%%%%%%%%%%%%%%%%
\label{sec:Intro}
The QCD Lagrangian with $N_f$ massless quarks
is invariant under 
$SU(N_f)_L \times SU(N_f)_R \times U(1)_V \times U(1)_A$ 
chiral rotations.
This symmetry, however, is broken in two different ways:
$SU(N_f)_L \times SU(N_f)_R$ part is spontaneously 
broken to $SU(N_f)_V$ in the QCD vacuum, 
while $U(1)_A$ is broken explicitly at the quantum level by the anomaly.

At a finite temperature $T$, 
it is widely believed that the $SU(N_f)_L \times SU(N_f)_R$ chiral
symmetry is recovered above the (critical) temperature $T_c \sim 150$ MeV
and plenty of evidence has been reported in
the first principle calculations of lattice QCD. 
For the $U(1)_A$ part, however, it remains an open question 
{\it if, how} ({\it much}), {\it and when} the symmetry is restored.
We only know that the $U(1)_A$ symmetry should be recovered in  the $T\to \infty$ limit,
where fermions eventually decouple as the lowest Matsubara frequency goes to infinity, so that 
the anomaly term cannot survive.

In particular, the question of whether  the $U(1)_A$ symmetry is restored or not near $T_c$
is of phenomenological importance.
For simplicity, let us consider the $N_f=2$ case.
As Pisarski and Wilczek \cite{Pisarski:1983ms} have discussed, 
the order of the phase transition may depend on the fate of the $U(1)_A$ symmetry:
if it remains to be broken at $T_c$, 
the chiral phase transition can be the second order,
while it is likely to be the first order when the $U(1)_A$ symmetry is also restored.
Furthermore the particle spectrum with the presence or absence of the $U(1)_A$ symmetry is quite different\cite{Shuryak:1993ee}. 
A connection between the restoration of $U(1)_A$ symmetry and the gap in the eigenvalue density of the Dirac operator near the origin is also suggested\cite{Cohen:1997hz}.

In principle, the fate of the $U(1)_A$ symmetry and related issues can be investigated
by numerical lattice QCD simulations 
\cite{Bernard:1996iz, Chandrasekharan:1998yx}.
Such studies are, however, still not easy,  since
both chiral and thermodynamical 
(the infinite volume) limits are required.
Currently, four simulations with different quark actions are on-going, 
but they have reported different results. 
Two of them \cite{Ohno:2011yr, :2012ja} have reported  that the eigenvalue density
of the Dirac operator has no gap at the origin
and its quark mass scaling is consistent 
with the $U(1)_A$ broken scenario.
Another group \cite{Kovacs:2011km} has also reported
no gap at the origin but they have found that 
small eigenmodes, which mainly contribute to $U(1)_A$ breaking correlation functions, are localized and uncorrelated, suggesting  that their contribution to the correlation functions  is negligible.
A simulation with overlap quarks \cite{Cossu:2012gm}, however, has reported 
the existence of a gap in the Dirac eigenvalue density
and a degeneracy of pion and eta(-prime) meson correlators, which suggests the recovery of $U(1)_A$ symmetry.

In this work, we address these problems 
again on a lattice, but using an analytic method.
For simplicity, we concentrate on the $N_f=2$ case in this paper.
We employ the overlap Dirac operator 
\cite{Neuberger:1997fp, Neuberger:1998wv}, 
which ensures the exact
$SU(2)_L\times SU(2)_R$ symmetry \cite{Luscher:1998pqa}
through the Ginsparg-Wilson relation \cite{Ginsparg:1981bj}
but the $U(1)_A$ is (correctly) broken by the fermonic measure\cite{Hasenfratz:1998ri}.
By using the spectral decomposition of the
multi-point correlation functions, and
assuming the restoration of the $SU(2)_L\times SU(2)_R$ symmetry,
we investigate if there are new constraints on the Dirac eigenvalue density in addition to
the manifest one implied by the well-known Banks-Casher relation \cite{Banks:1979yr}.
We also investigate whether the effect of the $U(1)_A$ symmetry breaking  disappear or not above $T_c$.

Since similar analytical investigations have been made
in previous studies, let us here revisit them
and make clear what is new in this work.
The first analysis based only on QCD was done by Cohen \cite{Cohen:1996ng}.
Assuming an absence of the zero-mode's contribution,
they concluded that all the {\it disconnected} contributions of the two-point functions 
in the $SU(2)_L\times SU(2)_R$ symmetric phase
disappear in the chiral limit.
This means that the pion, sigma, delta, and eta(-prime) meson 
correlators are all identical, realizing the  $U(1)_A$ symmetry. 
In this work, we include the zero-mode contribution explicitly to check whether 
or not this conclusion survives.

In fact, Lee and Hatsuda \cite{Lee:1996zy} 
(see also a related work by Evans {\it et al.} \cite{Evans:1996wf})
claimed that the zero-mode's contribution does not vanish but
keeps the disconnected contribution  of the scalar channel non-zero:
\begin{eqnarray}
\label{eq:LeeHatsuda}
\lim_{m\to 0}\left(\langle \bar{q}(x)q(x)\; \bar{q}(y)q(y)\rangle
-\langle \bar{q}(x)T^3 q(x)\; \bar{q}(y)T^3 q(y)\rangle\right)
\hspace{-3in}\nonumber\\
&=& \lim_{m\to 0}
\frac{\displaystyle\int d[A]_{\nu =\pm 1}e^{-S_{YM}}{\rm det}^\prime [\Dslash+m]^2\times 4\bar{\phi^A_0}(x)\phi^A_0(x)\bar{\phi^A_0}(y)\phi^A_0(y)}{Z}
+O(m),
\end{eqnarray}
where $q$ denotes the quark field, $T^3$ is the 3rd generator
of $SU(2)$.
On the right-hand side (RHS), $Z$ is the partition function of QCD,
$d[A]_{\nu =\pm 1}$ denotes the gauge field integrals
with a fixed topological charge $\nu=\pm 1$,
$S_{YM}$ denotes the gauge part of the action,
${\rm det}^\prime [\Dslash+m]={\rm det}[\Dslash+m]/m$ is 
the (continuum) fermion determinant with the quark mass $m$
from which the zero-mode contribution is subtracted,
and $\phi^A_0$ is an eigenfunction for the zero-mode
at a given configuration $A$.

The thermodynamical limit of Eq.~(\ref{eq:LeeHatsuda}) is, 
however, non-trivial and subtle, as was pointed out 
by Cohen \cite{Cohen:1997hz}.
In fact, we find that the RHS of Eq.~(\ref{eq:LeeHatsuda}) is
at least an $O(1/V)$ quantity.
Integrating Eq.~(\ref{eq:LeeHatsuda}) 
over $x$ and then taking an average over $y$,
(which should be grater than the original LHS), 
one immediately obtains
\begin{eqnarray}
\frac{1}{V}\int d^4y \int d^4 x (\mbox{LHS of Eq.(\ref{eq:LeeHatsuda}))}
&=& C\frac{4 N_\phi^2}{V} \to 0, \quad V\rightarrow \infty ,
\end{eqnarray}
since both $N_\phi=\int d^4 x \bar{\phi}_0^A(x)\phi_0^A(x)$
and $C=\lim_{m\to 0}(Z_1+Z_{-1})/m^2Z$ (where $Z_{\pm 1}$ denotes
the partition function in the topological sector of $\nu=\pm 1$)
are finite.
In fact, our work will show that not only Eq.~(\ref{eq:LeeHatsuda}) 
but also any contributions from zero modes of the Dirac operator
are in general $O(1/V)$ quantities, 
and thus disappear in the large volume limit.
It is not difficult to intuitively understand our result.
In the large
volume limit that $V\to \infty$ , the number of the fermion modes contributing to
the denominator $Z$ increases (it is natural to assume it to be
proportional to V ) while that of the numerator, where the bulk $O(V)$ contributions are canceled, is fixed to be $O(1)$.

Two years later, Cohen \cite{Cohen:1997hz} discussed a constraint on the 
eigenvalue density of the QCD Dirac operator in the chiral limit.
Relating the scalar one-point function and pseudoscalar
two-point functions in the chiral symmetric phase, 
he concluded that the eigenvalue density near zero,
\begin{equation}
\rho(\lambda) \sim |\lambda|^{\alpha},
\end{equation} 
must have $\alpha>1$.
In this paper, we examine up to 4 point correlation functions
in more various channels, and obtain a stronger constraint: $\alpha > 2$.
In the case of integer $\alpha$,  we believe that our constraint that $\alpha$ is equal to or larger than 3 should be the {\it strongest},
since we know of a theory which has both $\alpha=3$ and unbroken $SU(2)_L\times SU(2)_R$ (and also $U(1)_A$) chiral symmetries:  2-flavor massless free quarks.

Although we perform no numerical analysis in our study,
we would like to discuss possible artifacts in lattice QCD simulations.
In our analysis, the fully recovered  $SU(2)_L\times SU(2)_R$ symmetry is crucial.
We discuss possible modifications to our conclusions due to discretization effects
if a non-chiral quark action is employed in numerical simulations.
We also comment on finite volume effects.

Our paper is organized as follows.
In Section \ref{sec:setup},
we explain our setup, what we observe, and what we assume.
The constraints on the eigenvalue density with integer power at the origin
are given in Section~\ref{sec:constraints}.
In Section~\ref{sec:singlet}, we address a question on the fate of the $U(1)_A$ symmetry.
In Section~\ref{sec:syserr}, we discuss possible systematic effects which may arise in lattice QCD simulations. 
Section~\ref{sec:fractional} is devoted to a case where the eigenvalue density has a fractional power at the origin.
A conclusion and discussion are given in Section~\ref{sec:conclusion}.
Some useful formula and detailed calculations are collected in two appendices.

%%%%%%%%%%%%%%%%%%%%%%%%%%%%%%%%%%%%%%
%%%%%%%%%%%%%%%%%%%%%%%%%%%%%%%%%%%%%%
\section{lattice setup}
%%%%%%%%%%%%%%%%%%%%%%%%%%%%%%%%%%%%%%
%%%%%%%%%%%%%%%%%%%%%%%%%%%%%%%%%%%%%%
\label{sec:setup}
%%%%%%%%%%%%%%%%%%%%%%%%%%%%%%%%%%%%%%
\subsection{Spectral decomposition of the overlap fermion}
%%%%%%%%%%%%%%%%%%%%%%%%%%%%%%%%%%%%%%

We consider $N_f$-flavor lattice QCD in a finite volume $V$,
with the (anti-)periodic boundary condition in space(time). 

The quark part of the action is given by
\beqa
\label{eq:overlapaction}
S_F &=& a^4 \sum_x \left[ \bar\psi D(A) \psi + m \bar\psi F( D(A)) \psi\right](x),  \quad F(D) = 1 - \frac{R a}{2} D,
\eeqa
where $\psi=(\psi_1,\psi_2,\cdots \psi_{N_f})^T$ denotes the set of
$N_f$ fermion fields with the degenerate mass $m$, 
$a$ is the lattice spacing, and 
$D(A)$ is the overlap Dirac operator \cite{Neuberger:1997fp, Neuberger:1998wv} 
for a given gauge field $A$,
\begin{eqnarray}
D(A) &=& \frac{1}{Ra}\left(1+
\frac{D_W(A)-1/Ra}{\sqrt{(D_W(A)-1/Ra)^\dagger (D_W(A)-1/Ra)}} \right).
\end{eqnarray}
Here $D_W(A)$ denotes the Wilson Dirac operator
for the same gauge configuration $A$,
and $R$ is an arbitrary constant.
We have omitted the identity matrix ${\bf 1}_{N_f \times N_f}$
for the flavor indices for simplicity.

It is well-known that the overlap Dirac operator 
satisfies the $\gamma_5$ hermiticity, $D(A)^\dagger = \gamma_5 D(A)\gamma_5$,
and the Ginsparg-Wilson (GW) relation
\cite{Ginsparg:1981bj},
\beqa
D(A)\gamma_5 + \gamma_5 D(A) = a D(A) R \gamma_5 D(A).
\label{eq:GW1}
\eeqa
With this relation, the action Eq.~(\ref{eq:overlapaction})
in the $m\to 0$ limit is exactly symmetric \cite{Luscher:1998pqa}
under the {\it lattice} chiral rotation,
\beqa
\label{eq:chiral}
\delta_a \psi (x) &=& i\theta T_a \gamma_5 [(1-R aD(A))\psi](x), \\
\delta_a \bar\psi (x) &=& i\theta \bar\psi(x) T_a\gamma_5,
\eeqa 
where $\theta$ is an infinitesimal real parameter,
and $T_a$ denotes the  generator of $SU(N_f)$ for $a=1,2,\cdots N_f^2-1$,
and $T_0(={\bf 1}_{N_f \times N_f})$ denotes that for $U(1)_A$. 

We now consider eigenvalues and eigenfunctions of $D(A)$: 
$D(A)\phi_n^A =\lambda_n^A \phi_n^A$. The GW relation implies that
\beqa
\lambda_n^A + \bar\lambda_n^A &=& a R \bar\lambda_n^A\lambda_n^A,
\label{eq:GW2}
\eeqa 
where $\lambda_n^A$, and its complex conjugate $ \bar\lambda_n^A$ are in general complex numbers 
and therefore $(\phi_n^A)^\dagger D(A)^\dagger =(\phi_n^A)^\dagger \bar\lambda_n^A$.
Moreover, from the GW relation (\ref{eq:GW1}) and its consequence (\ref{eq:GW2}) we have
\beqa
D(A)\gamma_5 \phi_n^A &=& -\frac{\lambda_n^A}{1-Ra\lambda_n^A}\gamma_5\phi_n^A = \bar\lambda_n^A\gamma_5\phi_n^A.
\eeqa 
Since
$(\bar\lambda_n^A-\lambda_m^A) (\phi_n^A,\gamma_5\phi_m^A) = 0$,
eigenfunctions with complex eigenvalues can be ortho-normalized as
$(\phi_n^A, \phi_m^A)=(\gamma_5\phi_n^A, \gamma_5\phi_m^A)=\delta_{nm}$, 
and $(\phi_n^A, \gamma_5\phi_m^A) = 0$.
Here an inner product is defined as 
$(f,g) \equiv a^4 \sum_x f^\dagger (x) g(x) $.
For the real eigen values $\lambda_k^A=0$ and $\lambda_K^A=2/(Ra)$, 
their eigenfunctions can be chiral eigenstates,
since $D(A)$ and $\gamma_5$ commute for these real modes.
In the following, let us denote the number of the
left(right)-handed zero eigenmodes as $N_{L}$($N_R$) 
and that of the left(right)-handed $\lambda^A_K=2/Ra$ 
 (doubler) eigenmodes as $n_{L}$($n_R$).

Thus the propagator of the massive overlap fermion (for each flavor)
can be expressed in terms of these eigenvalues and eigenfunctions as,
\beqa
\label{eq:spectrumS}
S_A(x,y) 
&=&\sum_{\{ n\, \vert\, {\rm Im} \lambda_n^A > 0\}} \left[\frac{\phi_n^A(x) \phi_n^A(y)^\dagger}{f_m \lambda_n^A + m} +
\frac{\gamma_5\phi_n^A(x) \phi_n^A(y)^\dagger\gamma_5}{f_m \bar\lambda_n^A + m}\right] \nn\\&+&
\sum_{k=1}^{N_{R+L}^A} \frac{\phi_k^A(x)\phi_k^A(y)^\dagger}{m} + \sum_{K=1}^{n_{R+L}^A} \frac{\phi_K^A(x)\phi_K^A(y)^\dagger}{2/(Ra)},
\eeqa
where $f_m = 1 - R ma/2$,
$N_{R+L}^A=N_R^A+N_L^A$ is the total number of zero-modes, and
$n_{R+L}^A=n_{R}^A+n_{L}^A$  is the total number of doubler modes.

A measure for a given gauge field $A$ can be also 
written in terms of eigenvalues as
\beqa
P_m(A) &=& e^{-S_{YM}(A)} m^{N_f N_{R+L}^A} (\Lambda_R)^{N_f n_{R+L}^A}\
\prod_{{\rm Im} \lambda_n > 0} (Z_m^2 \bar\lambda_n^A \lambda_n^A + m^2 )^{N_f},
\eeqa
where $S_{YM}(A)$ is the gauge part of the action
(whose explicit form is not needed in this work), 
$\Lambda_R=2/(Ra)$ and $Z_m^2 = 1-m^2/\Lambda_R^2$.
Note that for even $N_f$, $P_m(A)$ is positive definite 
and an even function of $m$.

It is important to note that all quantities which
consist of $S_A(x,y)$ and $P_m(A)$ are finite at $V < \infty$, $m\not=0$ and $a\not=0$.
We then  carefully take the $V\to \infty$ and $m\to 0$ limits, without worrying about possible ultra-violet(UV) divergences,
until we eventually take the continuum limit.
 
%%%%%%%%%%%%%%%%%%%%%%%%%%%%%%%%%%%%%%
\subsection{Chiral Ward-Takahashi identities on the lattice}
%%%%%%%%%%%%%%%%%%%%%%%%%%%%%%%%%%%%%%

Now let us study the quantum aspects of the symmetry, 
performing the functional integral 
of an operator ${\cal O}$ over the quark fields,
\begin{eqnarray}
\langle {\cal O}\rangle_F \equiv
\int d\psi d\bar{\psi} \;{\cal O}e^{-S_F}.
\end{eqnarray}
The global lattice chiral rotation Eq.~(\ref{eq:chiral}) gives
the integrated Ward-Takahashi (WT) identity,
\begin{eqnarray}
\left\langle (\delta_{a0}J_0-\delta_a S_F){\cal O}+\delta_a {\cal O} \right\rangle_F =0,
\end{eqnarray}
where $J_0$ is the contribution from the chiral anomaly, 
or the Jacobian of the measure,
\beqa
J_0&= &- 2i N_f a^4 \sum_x\, \sum_{N=n,k,K} \phi_N^A(x)^\dagger
\gamma_5 \left(1- \frac{R}{2} a D\right)\phi_N^A(x) \nonumber \\
&=& -2 i N_f\times Q(A),
\label{eq:index}
\eeqa
where $Q(A)=N^A_R - N^A_L$ is the index of the overlap Dirac operator \cite{Hasenfratz:1998ri}, 
which gives an appropriate definition of the topological charge for
the given gauge configuration $A$.

In this paper, we consider the (volume integrals of) 
scalar and pseudoscalar density operators
\beqa
\label{eq:S}
S_a &=& a^4\sum_x [\bar\psi T_a (F(D(A))\psi](x), \\
P_a &=& a^4 \sum_x [\bar\psi T_a i\gamma_5 (F(D(A))\psi](x), 
\label{eq:P}
\eeqa
and their correlations. These two operators are transformed as
\beqa
\delta_b S_a &=& 2 \sum_c d_{ab}^c P_c, \;\;\;\;\;
\delta_b P_a = -2 \sum_c d_{ab}^c S_c, 
\eeqa
where $\left\{T_a,T_b\right\} = 2 \sum_c d_{ab}^cT_c$. 
In particular, in the $N_f=2$ case, we have
\beqa
\delta_b S_a &=&2\delta_{ab} P_0, \quad 
\delta_b P_a = -2\delta_{ab} S_0,\quad
(\mbox{for } a,b =1,2,3), \\
\delta_0 S_a &=&\delta_a S_0 = 2 P_a, \quad
\delta_0 P_a =\delta_a P_0 = -2 S_a, \quad (\mbox{for } a=0,1,2,3),
\eeqa
where we have adopted the normalization $(T^a)^2=1_{2\times 2}$ without summation on $a$.
It is now obvious that our mass term in the action Eq.~(\ref{eq:overlapaction})
can be simply expressed by $m S_0$,
and its transformation is $\delta_a S_F= 2 m P_a$.

%%%%%%%%%%%%%%%%%%%%%%%%%%%%%%%%%%%%%%
\subsection{Basic properties and assumptions}
%%%%%%%%%%%%%%%%%%%%%%%%%%%%%%%%%%%%%%
In this subsection, we explicitly give 
the basic properties and assumptions used in this paper.

If the $SU(2)_L \times SU(2)_R$ chiral symmetry is restored at $T > T_c$, 
we should have
\beqa
\lim_{m\rightarrow 0} \lim_{V\rightarrow \infty} 
\langle \delta_a  {\cal O} \rangle_m &=& 0
\quad (\mbox{for } a\not=0),
\label{eq:chiral_restor}
\eeqa
for an arbitrary operator ${\cal O}$,
where an average over gauge fields is defined by
\beqa
\langle {\cal O}(A)\rangle_m &=&\frac{1}{Z} \int{\cal D}A\, P_m(A)\, {\cal O}(A),\quad
Z = \int {\cal D}A\, P_m(A). 
\eeqa
Here we have included the subscript $m$ to remind the readers of the 
$m$-dependence.

In the following analysis, we will normalize the operator ${\cal O}$
(by multiplying $1/V^k$ with an integer $k$ ) so that  
$\lim_{V\rightarrow\infty}\langle \delta_a {\cal O} \rangle$ is well-defined.
Note that  $P_m(A)$ is positive for even $N_f$  and  $\int {\cal D}A P_m(A)/Z =1$. 

In our analysis, we assume that the vacuum expectation values of the $m$-independent observable ${\cal O}(A)$ is an analytic function of $m^2$, if the chiral symmetry is restored.
Therefore
if ${\cal O}(A)$  is $m$-independent and positive for all $A$, and is shown to satisfy
\beqa
\lim_{m\rightarrow 0}\frac{1}{m^{k}} \langle {\cal O}(A)^{l_0} \rangle_m &=& 0
\eeqa
with a non-negative integer $k$ and a positive integer $l_0$, we can write
\beqa
\langle {\cal O}(A)^{l_0} \rangle_m &=&  m^{2([k/2]+1)} \int{\cal D}A \hat P(m^2, A) {\cal O}(A)^{l_0} ,
\label{eq:m-dep}
\eeqa
where $[c]$ is the largest integer not larger than $c$, $\hat P(0,A) \not=0$ for $^\exists A$ and $\displaystyle \int{\cal D}A \hat P(m^2, A) {\cal O}(A)^{l_0}$ is non-negative and assumed to be finite in the large volume limit. In other words, the leading $m$  dependence arises from the contribution of configurations which satisfy $ \hat P(0,A) \not=0$. 

Under the above assumption, it is easy to see that
\beqa
\langle {\cal O}(A)^l \rangle_m &=& m^{2([k/2]+1)}
\int {\cal D}A\, \hat P(m^2,A){\cal O}(A)^l =O(m^{2([k/2]+1)}), 
\label{eq:power}
\eeqa
for an arbitrary positive integer $l$, as long as   $\displaystyle \int{\cal D}A \hat P(m^2, A) {\cal O}(A)^{l}$
is finite,  
since ${\cal O}(A)^{l_0}$ and ${\cal O}(A)^l$ are both positive and therefore share the same support 
in the configuration space. 

More generally, if a set of non-negative $m$-independent  functions ${\cal O}_i(A)$ satisfies 
$
\langle {\cal O}_i(A)\rangle_m = O(m^{2n_i})
$
with non-negative integers $n_i$ ($i=1,2,3,\cdots k$), it is easy to see that
\beqa
\label{eq:powerassumption}
\left\langle \prod_i^k {\cal O}_i(A) \right\rangle_m = O( m^{2n_{\rm max}}),
\eeqa
where $n_{\rm max} = \max(n_1,n_2,\cdots, n_k)$.

If  a non-negative ${\cal O}_0$ and an arbitrary operator ${\cal O}_1$ are $m$-independent and satisfy  
\begin{eqnarray}
\left\langle {\cal O}_0\right\rangle_m &=&O(m^{2n_0}), \qquad
\left\langle {\cal O}_1\right\rangle_m =O(m^{2n_1}),
\end{eqnarray}
we then have
\begin{eqnarray} 
\left\langle {\cal O}_0{\cal O}_1\right\rangle_m &=&
m^{2n_0}\int {\cal DA}\, {\cal O}_0(A) {\cal O}_1(A) \left\{ \hat P_+(m^2,A) + \hat P_-(m^2,A)\right\} =
O(m^{2n_0}),
\label{eq:powerassumption2}
\end{eqnarray}
irrespective of values of $n_0$ and $n_1$, where $\hat P(m^2, A) = \hat P_+(m^2,A) + \hat P_-(m^2,A)$ and
\beqa
{\cal O}_1(A)  P(m^2,A) &=&\left\{
\begin{array}{lc}
{\cal O}_1(A) P_+(m^2,A),  & {\cal O}_1(A) > 0 \\
{\cal O}_1(A) P_-(m^2,A),   & {\cal O}_1(A) <   0 \\
0, & {\cal O}_1(A) = 0 \\
\end{array} 
\right.  .
\eeqa

As will be seen later, we have  
\beqa
\lim_{m\rightarrow 0}\lim_{V\rightarrow\infty}\frac{1}{m V}\langle N_{R+L}^A \rangle_m &=& 0 ,
\label{eq:ev_density}
\eeqa
as a constraint from the chiral symmetry restoration. This leads to
\beqa
\lim_{V\rightarrow\infty}\frac{1}{V}\langle N_{R+L}^A \rangle_m &=& O(m^2).
\eeqa
This condition is, however, much weaker than the naive expectation that
the configuration $A$, which gives $N_{R+L}^A=O(V)$ has the weight $P_m(A) \propto m^{N_f O(V)}$ and therefore is much more suppressed in the large volume limit.
We do not assume such a highly suppressed weight $P_m(A)$ in this paper.
As will be shown later, however,  we can further prove that 
\beqa
\lim_{V\rightarrow\infty}\frac{1}{V}\langle N_{R+L}^A \rangle_m &=& 0 ,
\eeqa
for small enough $m$, using our weaker assumption, Eq.~(\ref{eq:m-dep}).

Note that analyticity in $m^2$ for physical observables and its consequence Eq.~(\ref{eq:m-dep})
do not hold at $T < T_c$, where the chiral symmetry is spontaneously broken. For example, the topological charge $Q(A)$ is expected to satisfy
\beqa
\lim_{V\rightarrow\infty}\frac{1}{V}\langle Q(A)^2\rangle_m = 
\frac{m\Sigma}{N_f} +O(m^2) ,
\eeqa
where $\Sigma$ is the chiral condensate.
The odd power of $m$ reflects the non-analyticity of 
the QCD partition function at $m=0$.

In the following analysis, the thermodynamical limit of
the eigenvalue density for a given configuration $A$,
\beqa
\rho^A(\lambda) &=& \lim_{V\rightarrow\infty} \frac{1}{V}\sum_{n\;({\rm Im}\lambda_n^A >0)} \delta(\lambda -\sqrt{\bar\lambda_n^A\lambda_n^A}),
\eeqa
plays a crucial role.
Since the temperature of the system is fully controlled by $P_m(A)$,
the eigenvalue density $\rho^A(\lambda)$ itself is 
not sensitive to the temperature\footnote{
The gauge configuration average 
$\langle \rho^A(\lambda)\rangle_m$
does, of course, depend on the temperature.
}.
It is also notable that $\int_0^{\Lambda_R} d\lambda \;\rho(\lambda)$
is finite on the lattice.
Therefore, $\rho^A(\lambda)$ is positive semi-definite
for arbitrary choice of $\lambda$ and $A$.\footnote{Strictly speaking, $\rho^A(\lambda)$ has a logarithmic divergence in the continuum limit, which can be absorbed by multiplying the quark mass $m$.}

Although the original eigenvalue spectrum 
at finite $V$ is a sum of delta functions,
we expect that such a spiky feature is smeared out
in the thermodynamical limit, and 
$\rho^A(\lambda)$ becomes a smooth function.
We here further assume that $\rho^A(\lambda)$ can be analytically 
expanded around $\lambda=0$
\footnote{
More precisely we here assume that configurations which do not have the expansion (\ref{eq:EV_exp}) are measure zero in the path integral with $P_m(A)$. 
This assumption excludes a possibility that $\langle \rho^A(\lambda)\rangle_m$ has a fractional power such that $\langle \rho^A(\lambda)\rangle_m\sim \lambda^\gamma$  with non-integer $\gamma$ at small $\lambda$. 
We consider the fractional case later in Sec.~\ref{sec:fractional}. }:
\beqa
\rho^A(\lambda) &=& \sum_{n=0}^\infty\rho^A_n \frac{\lambda^n}{n!} .
\label{eq:EV_exp}
\eeqa
An arbitrarily small convergence radius of this expansion, denoted by $\epsilon$, works well for
our later discussion where we take the massless limit.
As is well-known as the Banks-Casher relation \cite{Banks:1979yr} and will be seen later, $\lim_{m\rightarrow 0}\langle \rho^A_0 \rangle_m \not=0$ means the spontaneous chiral symmetry breaking.

%%%%%%%%%%%%%%%%%%%%%%%%%%%%%%%%%%%%%%%%
%%%%%%%%%%%%%%%%%%%%%%%%%%%%%%%%%%%%%%%%
\section{Constraints from the $SU(2)_L\times SU(2)_R$ restoration}
%%%%%%%%%%%%%%%%%%%%%%%%%%%%%%%%%%%%%%%%
%%%%%%%%%%%%%%%%%%%%%%%%%%%%%%%%%%%%%%%%
\label{sec:constraints}
In the following analysis, we concentrate on the case with $N_f=2$.
In this section, we derive the constraints on the eigenvalue density 
of the Dirac operator in the $SU(2)_L\times SU(2)_R$ 
chiral symmetric phase at finite temperature.

%%%%%%%%%%%%%%%%%%%%%%%%%%%%%%%%%%%%%%%%
\subsection{WT identities for scalar and pseudo-scalar operators}

Let us consider a product of scalar and pseudoscalar operators
 defined in Eqs.~(\ref{eq:S}) and (\ref{eq:P}),
\beqa
{\cal O}_{n_1,n_2,n_3,n_4} = P_a^{n_1} S_a^{n_2} P_0^{n_3} S_0^{n_4},
\eeqa
where $a$ represents a non-singlet index
($a=1,2,3$).
Here and in the following, a summation over $a$ is not taken,
and we explicitly use ``$0$'' for
the singlet operators.

Non-trivial WT identities are obtained from the set
\beqa
\label{eq:Oa}
{\cal O}_a^{(N)} &\equiv& \{ {\cal O}_{n_1,n_2,n_3,n_4} \vert \, n_1+n_2={\rm odd},\ n_1+n_3={\rm odd},\
\sum_i n_i =N\},
\eeqa
which requires the operator to be a non-singlet, and parity odd.
More explicitly, we have at $T>T_c$
\beqa
\lim_{m\rightarrow 0} \lim_{V\rightarrow\infty}\frac{1}{V^k} \langle \delta_a{\cal O}_{n_1,n_2,n_3,n_4} \rangle_m &=& 0\;\;\;\mbox{for}\;\;\;{\cal O}_{n_1,n_2,n_3,n_4} \in {\cal O}_a^{(N)},
\eeqa
where 
\beqa
\delta_a{\cal O}_{n_1,n_2,n_3,n_4} &=& -2n_1 {\cal O}_{n_1-1,n_2,n_3,n_4+1}+2n_2  {\cal O}_{n_1,n_2-1,n_3+1,n_4} \nn \\
&-&2n_3 {\cal O}_{n_1,n_2+1,n_3-1,n_4}+2n_4  {\cal O}_{n_1+1,n_2,n_3,n_4-1} .
\eeqa
Note that the fermion integrals are performed before the
gauge integrals: $\langle {\cal O}\rangle_m = 
\langle \langle{\cal O}\rangle_F\rangle_m$ but
we have omitted $\langle\cdots\rangle_F$ for notational simplicity.
Here the minimum power $k$ which makes the $V\rightarrow \infty$ limit finite 
depends on the choice of ${\cal O}_{n_1,n_2,n_3,n_4}$.
For further details, such as a relation of $n_1,n_2,n_3,n_4$ 
to $k$, see Appendix~\ref{sec:volume}.

%%%%%%%%%%%%%%%%%%%%%%%%%%%%%%%%%%%%%%%%
\subsection{Constraints at $N=1$}
%%%%%%%%%%%%%%%%%%%%%%%%%%%%%%%%%%%%%%%%
At $N=1$, there is only one operator ${\cal O}_{1000}= P_a$ 
in  ${\cal O}_a^{(N=1)}$, which gives
\begin{equation}
\delta^a P_a = -2 S_0.
\end{equation}
Using the decomposition in Eq.~(\ref{eq:spectrumS}),
and the normalization conditions 
$(\phi_n^A, \phi_m^A)=(\gamma_5\phi_n^A, \gamma_5\phi_m^A)=\delta_{nm}$,  
the thermodynamical limit of the 
functional integral for $S_0$ is expressed as
\beqa
\lim_{V\rightarrow\infty}\frac{1}{V}\langle -S_0 \rangle_m 
&=&
 \lim_{V\rightarrow\infty}\frac{N_f}{V} \left\langle \frac{N_{R+L}^A}{m} 
+ \sum_{n\;({\rm Im}\lambda_n^A >0)} \frac{2m}{Z_m^2\bar\lambda_n^A\lambda_n^A+m^2}\left(1-\frac{\bar\lambda_n^A\lambda_n^A}{\Lambda_R^2}\right)\right\rangle_m
\nn \\
&=&
\lim_{V\rightarrow\infty}\frac{N_f}{mV} \langle N_{R+L}^A \rangle_m 
+N_f \langle I_1 \rangle_m,
\eeqa
where 
\begin{eqnarray}
I_1 &=& m \int_0^{\Lambda_R} d\lambda\; 
\rho^A(\lambda)\frac{2g_0\left(\lambda^2\right)}{Z_m^2\lambda^2+m^2},\qquad
g_0(x)=1-\frac{x}{\Lambda_R^2}.
\end{eqnarray}

In  the chiral limit $m\to 0$,
only the vicinity of $\lambda=0$  contributes to
the integral, since
\begin{eqnarray}
\int_\epsilon^{\Lambda_R} d\lambda\; 
\rho^A(\lambda)\frac{2g_0\left(\lambda^2\right)}{Z_m^2\lambda^2+m^2},
\end{eqnarray}
is finite for arbitrarily small but positive $\epsilon$,
and thus, does not contribute to $I_1$ in the limit.
Expanding $\rho^A(\lambda)$  for $\lambda< \epsilon$, (see Eq.~(\ref{eq:EV_exp})),
it is not difficult to obtain (see appendix \ref{sec:integral})
\begin{eqnarray}
I_1 &=& m \int_0^{\epsilon} d\lambda\; 
\rho^A_0\frac{2g_0\left(\lambda^2\right)}{Z_m^2\lambda^2+m^2}
 + O(m)
\nonumber\\
&=& \pi \rho^A_0 + O(m).
\label{eqn:I1}
\end{eqnarray}
 
As an exercise, let us consider the  $T < T_c$ case, where the chiral
symmetry is spontaneously broken. Assuming that $\lim_{V\rightarrow\infty}\langle N_{R+L}^A \rangle_m/V \to 0$
\footnote{
\label{footnote:ChPT}
In chiral perturbation theory, one can confirm that 
$\langle |Q(A)| \rangle_m/V$ is an $O(1/\sqrt{V})$ quantity 
(even when $m$ is finite). 
Since the minimum of $N_{R+L}^A$ is equal to $\vert Q(A)\vert$
in the topological sector of $Q(A)$,
it is natural to assume that
$\langle N_{R+L}^A \rangle_m/V$
is also $O(1/\sqrt{V})$.
Moreover, using the fact that there is no massless pole in the non-singlet scalar
correlator $\langle S_aS_a\rangle$ in the chiral limit,  one can show that $\langle N_{R+L}^A \rangle_m/V=O(m^2)$.
},
the famous Banks-Casher relation \cite{Banks:1979yr} is reproduced:
\beqa
\lim_{m\rightarrow 0}\lim_{V\rightarrow\infty}\frac{1}{N_f V}
\langle -S_0\rangle_m &=&
\pi \lim_{m\rightarrow 0} \langle\rho^A_0\rangle_m  
\left(= \pi \lim_{m\rightarrow 0} \langle\rho^A(0)\rangle_m\right)
\not= 0.
\label{eq:BC}
\eeqa

On the other hand, in the chiral symmetric phase $T>T_c$, we require
\beqa
\lim_{m\rightarrow 0}\lim_{V\rightarrow\infty}  \frac{1}{V}\langle -S_0\rangle_m &=& \lim_{m\rightarrow 0}\lim_{V\rightarrow\infty}\frac{N_f}{mV} 
\langle N_{R+L}^A \rangle_m 
+N_f \lim_{m\rightarrow 0} \langle I_1 \rangle_m = 0.
\eeqa
Since both $N_{R+L}^A$ and $I_1$ are positive,
it is equivalent to separately require
the following two constraints:
\beqa
\lim_{V\rightarrow\infty} \frac{N_f}{V}\langle N_{R+L}^A\rangle_m &=& O(m^2) ,
\quad 
\langle  \rho_0^A\rangle_m = O(m^2) .
\label{eqn:rho0}
\eeqa
Using Eqs.~(\ref{eq:powerassumption}) and (\ref{eq:powerassumption2}) 
$ \langle  \rho_0^A\rangle_m = O(m^2)$ implies 
$\langle I_1^2 \rangle_m=O(m^2)$,
which will be useful in the analysis below.

%%%%%%%%%%%%%%%%%%%%%%%%%%%%%%%%%%%%%%%%
\subsection{Contribution from zero modes at general $N$}
\label{sec:generalN}
%%%%%%%%%%%%%%%%%%%%%%%%%%%%%%%%%%%%%%%%
Before extending our analysis to higher $N$, 
let us discuss the fate of the zero-mode contribution at general $N$.
For this purpose we consider an operator ${\cal O}_{1,0,0,N-1}\in {\cal O}_a^{(N)}$, 
whose non-singlet chiral WT identity requires
\beqa
\lim_{m\to 0}\lim_{V\to \infty}\left(
-\langle {\cal O}_{0,0,0,N}\rangle_m 
+(N-1)\langle {\cal O}_{2,0,0,N-2}\rangle_m 
\right)=0. 
\eeqa

Its dominant contribution at large volume is
\beqa
-\frac{1}{V^N} \langle (S_0)^N \rangle_m &=& - N_f^N
\left \langle \left\{(-1) \left(\frac{N_{R+L}^A}{m V} + I_1\right)\right\}^N\right\rangle_m + O(V^{-1}), 
\eeqa
and, therefore, from  the positivity of $N_{R+L}^A$ and $I_1$,
\beqa
\lim_{V\rightarrow\infty}
\frac{\langle ( N_{R+L}^A )^N\rangle_m}{V^N} &=&\left\{
\begin{array}{ll}
 O(m^{N+2}) & (\mbox{for even } N ) \\
 O(m^{N+1}) & (\mbox{for odd } N) \\
\end{array}
\right. .
\eeqa

Since this holds for arbitrary $N$, 
and $N_{R+L}^A$ does not explicitly depend on $m$, 
we conclude that
\beqa
\lim_{V\rightarrow\infty} \frac{\langle  N_{R+L}^A \rangle_m}{V} = 0,
\eeqa
at small but non-zero $m$.

This result implies that any zero-mode's contributions 
to an arbitrary local operator are {\it measure-zero} in the thermodynamical limit,
as we have already seen an example in Section~\ref{sec:Intro} \cite{Lee:1996zy}.
Therefore, we hereafter set $\lim_{V\rightarrow\infty}\langle N_{R+L}^A\rangle_m /V = 0$ even at small but non-zero $m$.

%%%%%%%%%%%%%%%%%%%%%%%%%%%%%%%%%%%%%%%%
\subsection{Constraints at $N=2$}
%%%%%%%%%%%%%%%%%%%%%%%%%%%%%%%%%%%%%%%%
We next consider the $N=2$ case.
In this case, two WT identities from 
${\cal O}_{1001}$ and ${\cal O}_{0110} \in {\cal O}_a^{(N=1)}$
require that the so-called (non-singlet) chiral susceptibilities, 
\beqa
\chi^{\sigma-\pi} &=&\frac{1}{V^2}\langle S_0^2- P_a^2\rangle_m, \qquad
\chi^{\eta-\delta} =\frac{1}{V}\langle P_0^2- S_a^2\rangle_m
\eeqa
vanish in the $V\to \infty$ and $m\to 0$ limits at $T>T_c$.
The first one, $\chi^{\sigma-\pi}$, 
has already been examined in the previous subsection.
 
In a similar way to the $N=1$ case, 
$\chi^{\eta-\delta}$ can be expressed in terms of eigenvalues as
\beqa
\lim_{V\to\infty}\chi^{\eta-\delta} &=& \lim_{V\to\infty}\left\langle
-\frac{N_f^2}{m^2V} Q(A)^2\right\rangle_m 
+N_f\left\langle\left(\frac{I_1}{m}+I_2\right)\right\rangle_m,
\eeqa
where $I_2$ is defined by
\begin{eqnarray}
I_2 &=& 2\int_0^{\Lambda_R} d\lambda\; 
\rho^A(\lambda)
\frac{m^2g_0^2(\lambda^2)-\lambda^2g_0(\lambda^2)}
{(Z_m^2\lambda^2+m^2)^2}\nn\\
&=&\left(\frac{2}{\epsilon}+\frac{2\epsilon}{\Lambda_R^2}\right)\rho_0^A +\left(2+\frac{\epsilon^2}{\Lambda_R^2}-\log\frac{\epsilon^2}{m^2}\right)\rho_1^A+ O(1).
\end{eqnarray}
Noting that
\begin{eqnarray}
\frac{I_1}{m}+I_2 =2mI_3
= 4m^2 \int_0^{\epsilon} d\lambda\; \rho^A(\lambda)
\frac{g_0^2(\lambda^2)}{(Z_m^2\lambda^2+m^2)^2}
= \rho_0^A \frac{\pi}{m}+2\rho_1^A + O(m)
\label{eq:I1mI2}
\end{eqnarray}
for an arbitrarily small (positive) parameter $\epsilon$, 
and expanding $\rho^A(\lambda)$ around $\lambda=0$, 
we obtain a condition in the chiral limit that
\beqa
\lim_{m\to 0}\lim_{V\to\infty}\chi^{\eta-\delta} = 
N_f \lim_{m\rightarrow 0}\left[ 
-\lim_{V\rightarrow\infty}\frac{N_f \langle Q(A)^2 \rangle_m}{ m^2 V}
+\frac{\pi}{m}\langle\rho_0^A\rangle_m +2\langle\rho_1^A\rangle_m
+O(m) \right].
\label{eq:mI3}
\eeqa
Since we already know that $\langle \rho_0^A\rangle_m = O(m^2)$,
this condition leads to
\beqa
\lim_{V\rightarrow\infty}\frac{N_f \langle Q(A)^2 \rangle_m}{ m^2 V} &=& 2 \langle \rho_1^A\rangle_m + O(m^2).
\label{eq:rho1}
\eeqa
Note that there should be no $O(m)$ term in Eq.~(\ref{eq:rho1}) according to the analyticity in $m^2$.
Therefore the $O(m)$ term can not be canceled  in Eq.~(\ref{eq:mI3}) at non-zero $m$.

%%%%%%%%%%%%%%%%%%%%%%%%%%%%%%%%%%%%%%%%
\subsection{Constraints at $N=3$ }
%%%%%%%%%%%%%%%%%%%%%%%%%%%%%%%%%%%%%%%%
From the WT identities at $N=3$ except the one considered in the
subsection \ref{sec:generalN}, the four quantities,
\beqa
\chi_1 &=& \frac{\langle {\cal O}_{0201}\rangle_m}{V^2} 
= N_f^2 \left\langle  I_1I_2 \right\rangle_m + O(1/V),\nonumber\\
\chi_2 &=& \frac{\langle {\cal O}_{1110}\rangle_m}{V} =
-N_f\left\langle N_f \frac{Q(A)^2}{m^3V} - 2 I_3\right\rangle_m + O(1/V),\nonumber\\
\chi_3 &=& \frac{\langle {\cal O}_{0021}\rangle_m}{V^2} =  -N_f^2\left\langle   I_1\left( \frac{I_1}{m}-\frac{N_fQ(A)^2}{m^2 V}\right)  \right\rangle_m+ O(1/V),\nonumber\\
\chi_4 &=& \frac{\langle {\cal O}_{2001}\rangle_m}{V^2} =
 -\frac{N_f^2}{m} \left\langle  I_1^2  \right\rangle_m+ O(1/V),
\eeqa
should vanish after taking the $V\to \infty$ and $m\to 0$ limits.
Here $I_3$ (and its asymptotic form near the chiral limit) is given by
\begin{eqnarray}
I_3 &=& 2m\int_0^{\epsilon} d\lambda\; 
\rho^A(\lambda)\frac{g_0^2(\lambda^2)}{(Z_m^2\lambda^2+m^2)^2}
= 
\left(
\frac{\pi}{2m^2}-\frac{3\pi}{4\Lambda_R^2}\right)\rho_0^A
 +\frac{\rho_1^A}{m} + \frac{\pi}{4}\rho_2^A+O(m).
\end{eqnarray}

Substituting the explicit form of $I_1$ and $I_2$, the result
(\ref{eqn:rho0}) in the previous subsection and our assumptions in
Eq.~(\ref{eq:powerassumption}) and (\ref{eq:powerassumption2}) give
\begin{eqnarray}
\langle I_1 I_2\rangle_m &=&O(m), \qquad
\langle (I_1)^2 \rangle_m = O(m^2).
\end{eqnarray}
so that $\chi_1$ and $\chi_4$ automatically vanish
in the $V\to \infty$ and $m\to 0$ limits.
Using the same assumptions and the result Eq.~(\ref{eq:rho1}) the following
relations can also be shown
\beqa
\frac{\langle Q(A)^2 I_1 \rangle_m}{m^2 V} &=&
\pi \frac{ \langle Q(A)^2 \rho_0^A\rangle_m}{m^2 V}+O(m),\\ 
N_f \frac{\langle Q(A)^2\rangle_m}{m^3V} &=& 2  \frac{\langle \rho_1^A\rangle_m}{m} +O(m).
\eeqa

The two remaining non-trivial conditions are
\beqa
\lim_{m\rightarrow 0}\lim_{V\rightarrow\infty}\chi_2 &=& -\pi N_f \lim_{m\rightarrow 0}\left[\frac{ \langle  \rho_0^A\rangle_m}{m^2} + \frac{\langle \rho_2^A\rangle_m}{2}\right]=0,\\
\lim_{m\rightarrow 0}\lim_{V\rightarrow\infty}\chi_3 &=& \pi N_f^3\lim_{m\rightarrow 0}\lim_{V\rightarrow\infty}\frac{ \langle Q(A)^2\rho_0^A\rangle_m}{m^2 V}=0.
\eeqa
From the first condition, we obtain a constraint,
\beqa
\langle  \rho_0^A\rangle_m&=& - m^2 \frac{\langle \rho_2^A\rangle_m}{2} +O(m^4).
\label{eq:rho2}
\eeqa
Moreover, since $ \langle  \rho_0^A\rangle_m$ is positive 
(which is required by the positivity of $\langle \rho^A(0)\rangle_m$),
$ \langle \rho_2^A\rangle_m$ must be negative for small $m$.

The condition for $\chi_3$ leads to
\beqa
\lim_{V\rightarrow\infty}\frac{ \langle Q(A)^2 \rho_0^A\rangle_m}{m^2 V}  =O(m^2) .
\label{eq:Q-rho0}
\eeqa
This condition does not necessarily give stronger constraint 
than $\langle Q(A)^2\rangle_m = O(m^2V)$ and 
$\langle \rho_0^A\rangle_m=O(m^2)$, since it only requires that  
a set of gauge configurations that satisfies both $Q(A)^2\not=0$ and $\rho^A_0\not=0$ has a weight $m^4\hat P(A,m^2) +O(m^6)$.

%%%%%%%%%%%%%%%%%%%%%%%%%%%%%%%%%%%%%%%%
\subsection{Constraints at $N=4$ }
%%%%%%%%%%%%%%%%%%%%%%%%%%%%%%%%%%%%%%%%
The 8 WT identities at $N=4$ give 7 independent constraints 
\begin{eqnarray}
&&
\langle {\cal O}_{4000} -{\cal O}_{0004}\rangle_m \rightarrow 0, \quad
\langle {\cal O}_{4000} -3 {\cal O}_{2002}\rangle_m \rightarrow 0, \nonumber\\
&&
\langle {\cal O}_{0400} - {\cal O}_{0040}\rangle_m \rightarrow 0, \quad
\langle {\cal O}_{0400} -3 {\cal O}_{0220}\rangle_m \rightarrow 0, \nonumber\\
&&
\langle {\cal O}_{2020} -{\cal O}_{0202}\rangle_m \rightarrow0, \quad
\langle {\cal O}_{2200} - {\cal O}_{0022}\rangle_m \rightarrow 0, \nonumber\\
&&
\langle 2{\cal O}_{1111} - {\cal O}_{0202}+{\cal O}_{0022}\rangle_m \rightarrow 0 ,
\end{eqnarray}
where the $V\to\infty$ and $m\to 0$ limits are abbreviated by the arrows.
The $O(V^4)$ contribution from $S_0^4$ in the first equation has already
been considered.

At $O(V^3)$, there are 3 conditions,
\beqa
\lim_{V\to \infty}\frac{1}{V^3}\langle P_a^2S_0^2\rangle_m
&=&N_f^3\frac{\langle I_1^3\rangle_m}{m}\rightarrow 0, \\
\lim_{V\to \infty}\frac{1}{V^3}\langle S_a^2S_0^2\rangle_m
&=&-N_f^3\langle I_1^2 I_2\rangle_m\rightarrow 0,\\ 
\lim_{V\to \infty}\frac{1}{V^3}\langle P_0^2S_0^2\rangle_m
&=&N_f^3\left\langle I_1^2\left\{\frac{I_1}{m} 
-\frac{N_f Q(A)^2}{m^2V}\right\}\right\rangle_m\rightarrow 0.
\eeqa
It is not difficult to confirm that all of them
are automatically satisfied, since
\beqa
\langle I_1^n\rangle_m = \langle\left\{\pi\rho_0^A + O(m)\right\}^n\rangle_m = O(m^2),
\eeqa
for any integer $n \ge 2 $ from $\langle\rho_0^A\rangle_m =O(m^2)$, and 
\beqa
\frac{\langle I_1^2 Q(A)^2\rangle_m}{m^2V} = \frac{\langle \{\pi\rho_0^A+O(m)\}^2 Q(A)^2\rangle_m}{m^2V} = O(m^2),
\eeqa
from Eq.~(\ref{eq:Q-rho0})  together with the assumption
(\ref{eq:powerassumption2}) for remaining cross terms.
Namely, these three give no additional constraint.

At $O(V^2)$ we have
\beqa
\frac{1}{V^2}\langle S_a^4-P_0^4\rangle_m \rightarrow 0, \quad
\frac{1}{V^2}\langle S_a^4-3S_a^2P_0^2\rangle_m \rightarrow 0, \quad
\frac{1}{V^2}\langle P_a^2(P_0^2-S_a^2)-2P_aS_aP_0S_0\rangle_m \rightarrow 0.\nn \\
\label{eq:V2}
\eeqa
After a little algebra using the formulas 
in appendices~\ref{sec:trace} and \ref{sec:volume}, 
the first condition becomes
\beqa
3N_f^2\langle (I_2+I_1/m)(I_2-I_1/m) \rangle_m + \frac{6N_f^3}{m^3V}\langle Q(A)^2 I_1\rangle_m -\frac{N_f^4}{m^4V^2}
\langle Q(A)^4\rangle_m \rightarrow 0.
\label{eq:V2-1}
\eeqa
Using 
\beqa
I_2-\frac{I_1}{m} &=&
\rho_0^A\left(-\frac{\pi}{m}+\frac{4}{\epsilon}+\frac{4\epsilon}{\Lambda_R^2}\right)+\rho_1^A\left(2+\frac{2\epsilon^2}{\Lambda_R^2}-4\log\frac{\epsilon}{m}\right) + O(1) ,
\eeqa
Eq.~(\ref{eq:I1mI2}) and $\langle (\rho_0^A)^n\rangle_m =O(m^2)$, 
we can show that the first term in Eq.~(\ref{eq:V2-1}) is 
at most logarithmically divergent
in the limit $m\rightarrow 0$. 
Note that the second term is also logarithmically divergent due to cross
contributions from the $O(m)$ terms in $I_1$ (appendix \ref{sec:integral}) and
Eq.~(\ref{eq:rho1}).
Therefore, in order to satisfy Eq.~(\ref{eq:V2-1}), the last term should
not be power divergent and should at least fulfill
\beqa
\lim_{V\rightarrow\infty}\frac{1}{V^2}\langle Q(A)^4\rangle_m = O(m^4), 
\eeqa
 which leads to
 \beqa
 \lim_{V\rightarrow\infty}\frac{1}{V^k}\langle Q(A)^{2k}\rangle_m = O(m^4), 
 \eeqa
 for an arbitrary positive integer $k$. Combining this with Eq.~(\ref{eq:rho1}), 
we obtain a constraint on the spectral density,
 \beqa
 \langle \rho_1^A\rangle_m = O(m^2),
 \label{eq:rho1m2}
 \eeqa
so that Eq.~(\ref{eq:V2-1}) now becomes
\beqa
-3N_f^2 \frac{\pi^2}{m^2} \langle (\rho_0^A)^2\rangle_m -\frac{N_f^4}{m^4V^2}
\langle Q(A)^4\rangle_m \rightarrow 0.
\eeqa
Since both terms are negative semi-definite, this WT identity requires
\beqa
\langle (\rho_0^A)^k\rangle_m =O(m^4) , \quad 
\lim_{V\rightarrow\infty}\frac{1}{V^l}\langle Q(A)^{2l}\rangle_m = O(m^6), 
\eeqa
for arbitrary positive integers $k$ and $l$. The first condition also gives
\beqa
\label{eq:rho2cond}
\langle \rho_2^A\rangle_m =O(m^2),
\eeqa
from Eq.~(\ref{eq:rho2}). 

The last constraint Eq.~(\ref{eq:rho2cond}) can be obtained through a different argument. 
From Eq.~(\ref{eq:rho1m2}), the eigenvalues density near the chiral limit becomes 
\beqa
\langle \rho^A(\lambda)\rangle_m =  \langle \rho_2^A\rangle_m \frac{\lambda^2}{2} + O(\lambda^3)+O( m^2) .
\eeqa
The positivity of $\langle \rho^A(\lambda)\rangle_m$ implies
$ \langle \rho_2^A\rangle_m \ge 0 $ near $m=0$
but this contradicts with the positivity of $\langle \rho_0^A\rangle_m$ in
Eq.~(\ref{eq:rho2}), unless $\langle \rho_2^A\rangle_m =O(m^2)$,
and thus, $\langle \rho_0^A\rangle_m =O(m^4)$.

It is now easy to see that the second and third conditions in Eq.~(\ref{eq:V2}) are automatically satisfied:
the second one gives
\beqa
6N_f^2m\langle I_2I_3\rangle_m -\frac{3N_f^3}{m^2V}\langle I_2 Q(A)^2\rangle_m = O(m^2) + O(m^4),
\eeqa 
while the third one is evaluated as
\beqa
6N_f^2\left\langle I_1I_3\right\rangle_m -\frac{3N_f^3}{m^3V}\langle I_1 Q(A)^2\rangle_m = O(m^2) + O(m^3).
\eeqa

%%%%%%%%%%%%%%%%%%%%%%%%%%%%%%%%%%%%%%%%
\subsection{Special constraints at general $N$}
%%%%%%%%%%%%%%%%%%%%%%%%%%%%%%%%%%%%%%%%
\label{subsec:special}
In this subsection, we consider a special type of 
operators : ${\cal O}_{0,1,(4k-1),0}\in {\cal O}_a^{(N=4k)} $
at a general positive integer $k$, 
whose non-singlet WT identity gives a condition,
\beqa
\lim_{m\rightarrow 0} \langle  (4k-1) S_a^2 P_0^{4k-2} -P_0^{4k}\rangle_m  &=& 0.
\eeqa
At the leading order of $V$ ($V^{2k}$ in this case), the above 
condition corresponds to
\beqa
-(4k-1) \frac{N_f}{ V^{2k-1}} \langle I_2 P_0^{4k-2} \rangle_m - \frac{1}{V^{2k}}\langle P_0^{4k}\rangle_m &\rightarrow& 0,
\eeqa
in the chiral limit.
From the results in appendices~\ref{sec:trace} and \ref{sec:volume},
\beqa
\frac{\langle P_0^{4k}\rangle_F}{V^{2k}} &=& \sum_{n=0}^{2k} {}_{4k}C_{2n} (2n-1)!! \left(\frac{ -N_f^2 Q(A)^2}{m^2 V}\right)^{2k-n}\left(\frac{N_f I_1}{m}\right)^n +O(V^{-1}),
\eeqa
where we have used the definition $(-1)!!=1$.
The non-singlet WT identity is expressed by
\beqa
-(4k-1)\sum_{n=0}^{2k-1} {}_{4k-2}C_{2n} (2n-1)!! \left\langle  \left(\frac{ -N_f^2 Q(A)^2}{m^2 V}\right)^{2k-1-n}\left(\frac{N_f I_1}{m}\right)^nN_f I_2 \right\rangle_m\nn \\
-\sum_{n=0}^{2k} {}_{4k}C_{2n} (2n-1)!! \left\langle \left(\frac{ -N_f^2 Q(A)^2}{m^2 V}\right)^{2k-n}\left(\frac{N_f I_1}{m}\right)^n\right\rangle_m
\rightarrow 0.
\eeqa

From the above condition(s), we would like to inductively prove that
\begin{eqnarray}
\label{eq:Qcond}
\frac{\langle Q^{2l}\rangle_m}{V^{l}} = O(m^{4k+2}),
\quad
\langle (\rho_0^A)^l\rangle_m =O(m^{2k+2}),
\end{eqnarray}
holds for arbitrary positive integers $l$ and $k$.

Suppose
\begin{eqnarray}
\frac{\langle Q^{2l}\rangle_m}{V^{l}} = O(m^{4k-2}),\quad
\langle (\rho_0^A)^l\rangle_m =O(m^{2k}),
\label{eq:assume}
\end{eqnarray}
is obtained 
from the WTI at $N=4k-4$ (this is true for $k=2$).
The constraint above is then reduced to 
\beqa
\label{eq:induction}
&-&\left\langle (4k-1)!! \left(\frac{N_f I_1}{m}\right)^{2k-1} N_f (I_2+I_1/m) +\left(\frac{ -N_f^2 Q(A)^2}{m^2 V}\right)^{2k}
\right.\nonumber\\&&\left.
+(4k-1) \left(\frac{ -N_f^2 Q(A)^2}{m^2 V}\right)^{2k-1} N_f ( I_2 +2k I_1/m)\right\rangle_m
\rightarrow 0, 
\eeqa
where only those terms with $n=0, 2k-1$ in the first summation,
$n=0,1,2k$ in the second summation remain.
While the first and the third terms are finite and linearly divergent in the $m\rightarrow 0$ limit,  the second term is seen to be quadratically divergent  from  eq.~(\ref{eq:assume}) as
\beqa
\label{eq:induction2}
-\frac{N_f^{4k} }{m^{4k}}\frac{\langle Q(A)^{4k}\rangle_m}{V^{2k}} =O\left( m^{-2} \right).
\eeqa
In order for the WTI to be satisfied, the quadratic divergence should be
absent, so that $\langle Q(A)^{2l}/V^l \rangle_m=O(m^{4k})$
for an arbitrary positive integer $l$, thanks to 
Eq.~(\ref{eq:power}).

Using this result the third term disappears faster than the others
and the WTI becomes
\begin{eqnarray}
-(4k-1)!! N_f^{2k-1} \frac{\langle (\pi\rho_0^A)^{2k}\rangle_m}{m^{2k}} 
-\frac{N_f^{4k} }{m^{4k}}\frac{\langle Q(A)^{4k}\rangle_m}{V^{2k}}
\to0.
\end{eqnarray}
Note here that the both terms are negative semi-definite
and therefore each term {\it must vanish} in the chiral limit.
This completes the proof for Eq.~(\ref{eq:Qcond}).

Since $k$ can be arbitrarily large, 
 we now have another stronger constraint on the zero-mode's contribution:
\beqa
\label{eq:Q2cond}
\lim_{V\rightarrow\infty} \frac{\langle Q(A)^2\rangle_m}{V} &=& 0,
\eeqa
and that on the spectral density,
\begin{equation}
\langle \rho_0^A \rangle_m = 0,
\end{equation}
which hold even at small but non-zero $m$.

%%%%%%%%%%%%%%%%%%%%%%%%%%%%%%%%%%%%%%%%
\subsection{Short summary of the constraints}
%%%%%%%%%%%%%%%%%%%%%%%%%%%%%%%%%%%%%%%%
Here we summarize the constraints obtained in this section.
For the eigenvalue density, we have 
 \beqa
 \langle \rho_0^A\rangle_m =0, \quad
 \langle \rho_1^A\rangle_m = O(m^2), \quad
 \langle \rho_2^A\rangle_m = O(m^2),
 \eeqa
 at a small but non-zero $m$.
Namely, the eigenvalue density must have the form
 \beqa
\label{eq:resultrho}
\lim_{m\rightarrow 0} \langle \rho^A(\lambda)\rangle_m &=& \langle \rho_3^A\rangle_0 \frac{\lambda^3}{3!}+O(\lambda^4). 
 \eeqa
We believe that this new condition is not only stronger
than those found in previous works, 
but also the {\it strongest} since we know 
that the $N_f=2$ massless free quark theory
has $\langle \rho^A_3 \rangle_0 \neq 0$ 
keeping the exact chiral $SU(2)_L\times SU(2)_R$ (and $U(1)_A$) symmetry.
Therefore, it is very likely that we will not find any additional information
from $N\geq 5$ correlation functions.
%Note here that the density of eigenvalues is always 
%defined only in the $V\rightarrow\infty$ limit.
 
For the discrete zero modes, we have obtained
\beqa
\label{eq:resultzero}
\lim_{V\rightarrow \infty} \frac{1}{V^k} \langle (N_{R+L}^A)^k\rangle_m &=& 0,\quad
\lim_{V\rightarrow \infty} \frac{1}{V^k} \langle Q(A)^{2k}\rangle_m = 0,
\eeqa
for an arbitrary positive integer $k$ at a small but non-zero $m$.
These zero-modes give
no contribution to the correlation functions we are considering.

%%%%%%%%%%%%%%%%%%%%%%%%%%%%%%%%%%%%%%%%%
%%%%%%%%%%%%%%%%%%%%%%%%%%%%%%%%%%%%%%%%%
\section{Fate of $U(1)_A $ anomaly}
\label{sec:singlet}
%%%%%%%%%%%%%%%%%%%%%%%%%%%%%%%%%%%%%%%%%
%%%%%%%%%%%%%%%%%%%%%%%%%%%%%%%%%%%%%%%%%

In this section, we discuss how the constraints 
in the $SU(2)_L\times SU(2)_R$ symmetric phase, obtained
in the previous section, 
affect the $U(1)_A$ breaking correlators.
Here we consider a set of (pseudoscalar singlet) operators,
\beqa
{\cal O}_0^{(N)} &\equiv& \{ {\cal O}_{n_1,n_2,n_3,n_4} \vert \, n_1+n_2={\rm even},
\ n_1+n_3={\rm odd},\
\sum_i n_i =N\},
\eeqa
and its chiral $U(1)_A$ rotation,
\beqa
\delta_0{\cal O}_{n_1,n_2,n_3,n_4} &=& -2n_1 {\cal O}_{n_1-1,n_2+1,n_3,n_4}+2n_2  {\cal O}_{n_1+1,n_2-1,n_3,n_4} \nn \\
&-&2n_3 {\cal O}_{n_1,n_2,n_3-1,n_4+1}+2n_4  {\cal O}_{n_1,n_2,n_3+1,n_4-1} . 
\eeqa
For later convenience, let us also define a set of (scalar singlet) operators,
\beqa
{\cal O}^{(N)} &\equiv& \{ {\cal O}_{n_1,n_2,n_3,n_4} \vert \, n_1+n_2={\rm even},
\ n_1+n_3={\rm even},\
\sum_i n_i =N\}.
\eeqa
As QCD keeps the vector like $SU(2)_V$ symmetry and the parity symmetry,
any operator with a non-zero expectation value must be a member of 
${\cal O}^{(N)}$. Note that we have already introduced 
the set of pseudoscalar non-singlet operators 
${\cal O}_a^{(N)}$ in Eq.~(\ref{eq:Oa}).

Since the $U(1)_A$ transformation is anomalous, 
the expectation value of the variation 
$\langle \delta_0{\cal O} \rangle_m$ 
is nonzero [as shown by the WTI; see Eq.~(\ref{eq:index})],
\begin{eqnarray}
\label{eq:singletWT}
\lim_{m\to 0}\langle \delta_0{\cal O} \rangle_m 
&=& 2i N_f\lim_{m\to 0}\langle Q(A) {\cal O}\rangle_m,
\end{eqnarray}
and the $U(1)_A$ symmetry is broken.

It is, however, still possible to have zeros on the both sides
of Eq.~(\ref{eq:singletWT}). 
If this is the case, the $U(1)_A$ anomaly is invisible.
In fact, we show below that the constraints obtained in the previous section
are strong enough to suppress the variation 
$\langle \delta_0{\cal O} \rangle_m$ for ${\cal O}\in {\cal O}_0^{(N)}$,
to be zero
in the large volume $V\to \infty$ and chiral $m\to 0$ limits.
Namely, the $U(1)_A$ symmetry must be restored, at least, 
for the operator set ${\cal O}_0^{(N)}$.

%%%%%%%%%%%%%%%%%%%%%%%%%%%
\subsection{Odd $N$ case}
%%%%%%%%%%%%%%%%%%%%%%%%%%%

For the odd $N$ cases, we can show a relation for the number of
operators
$\vert {\cal O}_a^{(N)} \vert=\vert{\cal O}_0^{(N)} \vert
=\vert{\cal O}^{(N)}\vert$, where $\vert {\cal A}\vert$ denotes 
the number of independent operators in ${\cal A}$.
See appendix~\ref{sec:general} for the details.

As the exact chiral $SU(2)_L\times SU(2)_R$ symmetry
requires $\vert{\cal O}_a^{(N)}\vert$ independent
WT identities in the chiral limit to be zero, 
\begin{eqnarray}
\lim_{m\to 0}\langle \delta_a{\cal O}_i\rangle_m
=
\lim_{m\to 0}\sum_{j=1}^{\vert{\cal O}^{(N)}\vert}M_{ij}
\langle {\cal O}_j\rangle_m=0,\quad
{\cal O}_i\in {\cal O}_a^{(N)},\;
{\cal O}_j\in {\cal O}^{(N)},
\end{eqnarray}
where the matrix $M$ is specified by the WTI that one considers.
Since the chiral transformation keeps the independence of the operator,
it can be proved that $\det M\neq0$\footnote
{If $\det M=0$, we can construct a chiral invariant operator from a linear combination of operators in ${\cal O}_a^{(N)}$. Since all operators in ${\cal O}_a^{(N)}$ have odd numbers of the index $a$, however,  no chiral invariant operator should exist in ${\cal O}_a^{(N)}$.},
and the WTI requires
\begin{eqnarray}
\lim_{m\to 0}\langle {\cal O}_i\rangle_m=0\;\;\;\mbox{for any}\;
{\cal O}_i\in {\cal O}^{(N)},
\end{eqnarray}
or equivalently, that there is no operator 
in ${\cal O}^{(N)}, \delta_a{\cal O}_a^{(N)}$, and $\delta_0{\cal O}_0^{(N)}$
which has a non-zero expectation value 
in the $SU(2)_L\times SU(2)_R$ symmetric phase.

Since the $U(1)_A$ variation of any operator in  ${\cal O}_0^{(N)}$ is 
an element of ${\cal O}^{(N)}$, we can conclude that
\begin{equation}
\lim_{m\to 0}\langle \delta_0{\cal O} \rangle_m = 0\;\;\;\mbox{for any}\;
{\cal O}\in {\cal O}_0^{(N)}.
\end{equation}
Without referring any specific constraints obtained in the previous section,
we can thus show that the $U(1)_A$ breaking is invisible for these operators.

%%%%%%%%%%%%%%%%%%%%%%%%%%%
\subsection{$N=2, 4, $ and $6$}
%%%%%%%%%%%%%%%%%%%%%%%%%%%
For even $N$, the situation 
is not so simple as for the odd $N$'s 
(see Appendix~\ref{sec:general} for the details.).
We need to examine the WT identities explicitly.

At $N=2$, there remains one non-trivial susceptibility,
but one can immediately show that it should vanish:
%%%%%%%%%%%%%%%%%%%%%%%%%%%
\beqa
\label{eq:pieta}
\chi^{\pi-\eta} &=&\frac{1}{V}\langle P_a^2 - P_0^2 \rangle_m =
\lim_{V\rightarrow \infty} \frac{N_f^2}{m^2V}\langle Q(A)^2\rangle_m = 0, 
\eeqa
for small but non-zero $m$, 
thanks to Eq.~(\ref{eq:Q2cond}). 
Noting that $P_a^2 - P_0^2 = (P_a^2 -S_0^2) + (S_0^2-S_a^2)+(S_a^2- P_0^2)$,
we can also show that
\begin{eqnarray}
\label{eq:sigma-delta}
\chi^{\delta-\sigma} &=&\frac{1}{V^2}\langle S_a^2-S_0^2 \rangle_m = O(1/V)
+O(m^2).
\end{eqnarray}
Since LHS of Eq.~(\ref{eq:sigma-delta})
is the (double) volume average of LHS of Eq.~(\ref{eq:LeeHatsuda}),
this give another proof that the $U(1)_A$ breaking effect
in Ref.~\cite{Lee:1996zy}
cannot survive in the thermodynamical limit.

At $N=4$, there are two non-trivial susceptibilities, 
\beqa
\chi_5 &=& \langle {\cal O}_{0022} - {\cal O}_{2002}\rangle_m, \quad 
\chi_6 = \langle {\cal O}_{0022} - {\cal O}_{0220}\rangle_m.
\eeqa 
Neglecting $N_{R+L}^A /V$ and $Q(A)^2/V$ terms 
and using the constraint on $I_1$ obtained in the previous section,
both of them disappear  as
\beqa
\lim_{m\rightarrow 0}\lim_{V\rightarrow\infty}\frac{\chi_5}{V^3} &=& -\lim_{m\rightarrow 0}\lim_{V\rightarrow\infty}
N_f^3 \left\langle \frac{N_fQ(A)^2}{m^2 V} \left(\frac{N_{R+L}^A}{mV} + I_1\right)^2
\right\rangle_m = 0, \\
\lim_{m\rightarrow 0}\lim_{V\rightarrow\infty}\frac{\chi_6}{V^3} &=& \lim_{m\rightarrow 0}\lim_{V\rightarrow\infty} \frac{N_f^3} {m}  \left\langle \left(\frac{N_{R+L}^A}{mV} + I_1\right)^2\left(\frac{N_{R+L}^A}{mV} + I_1 -\frac{N_fQ(A)^2}{mV}\right) \right\rangle_m \nn \\
&=& \lim_{m\rightarrow 0}  \frac{N_f^3} {m }  \left\langle  I_1^3\right\rangle_m =0.
\eeqa

At $N=6$, we have 4 non-trivial susceptibilities, 
\beqa
\chi_7 &=&\langle {\cal O}_{0024}- {\cal O}_{2004}\rangle_m,\quad
\chi_8 =\langle {\cal O}_{2040}- {\cal O}_{0204}\rangle_m, \\
\chi_{9} &=&\langle {\cal O}_{0420}- {\cal O}_{0042}\rangle_m,\quad
\chi_{10} =\langle {\cal O}_{0042}- {\cal O}_{0024}\rangle_m.
\eeqa
In the large volume limit, they behave as
\beqa
\lim_{V\rightarrow\infty}\frac{\chi_7}{V^5} &=& 0,  \\
 \lim_{V\rightarrow\infty}\frac{\chi_{8}}{V^5} &=& N_f^5\langle I_2 I_1^4\rangle_m =O(m^4),\\
 \lim_{V\rightarrow\infty}\frac{\chi_{9}}{V^4} &=& -\frac{N_f^4}{m^2}\langle I_1^4\rangle_m =O(m^2), \\
 \lim_{V\rightarrow\infty}\frac{\chi_{10}}{V^5} &=& -\frac{N_f^5}{m}\langle I_1^5\rangle_m =O(m^4),
\eeqa
all of which vanish after the chiral limit is taken.

We thus conclude that the $U(1)_A$ symmetry breaking
is not viable for at least $N\leq 6$.

%%%%%%%%%%%%%%%%%%%%%%%%%%%
\subsection{General even $N$}
%%%%%%%%%%%%%%%%%%%%%%%%%%%
\label{subsec:U(1)generalN}

In order to consider the general $N$ case, 
let us look at RHS of Eq.~(\ref{eq:singletWT}).
Namely, if we can show that
\beqa
\lim_{m\rightarrow 0}\lim_{V\rightarrow\infty}\frac{1}{V^k}\langle Q(A) {\cal O} \rangle_m = 0,
\eeqa
with some appropriate power of $k$,
we can prove that LHS of Eq.~(\ref{eq:singletWT}) also vanishes.
In the analysis below, we divide ${\cal O}_0^{N}$ into two classes :
the one  with $(n_1, n_2, n_3)$=(even, even, odd), and
another with  $(n_1, n_2, n_3)$=(odd, odd, even).

For the former class, or more explicitly in the case of
$(n_1, n_2, n_3)=(2k_1, 2k_2, 2k_3+1)$, 
the leading contribution in 
$\delta_0 {\cal O}_{n_1,n_2,n_3,n_4}$ comes from 
$-n_3 {\cal O}_{n_1,n_2,n_3-1,n_4+1}$, whose leading contribution 
in $V$ has $O( V^k)$ with
$k=k_1+k_2+k_3+ n_4+1$ (see appendix~\ref{sec:volume}).
Therefore, we have
\begin{eqnarray}
\frac{iN_f}{V^k} \langle Q(A) {\cal O}_{n_1,n_2,n_3,n_4} \rangle_m
&\simeq&
N_f^{k+1} n_3 (2k_1-1)!!(2k_2-1)!!(2k_3-1)!! \nn \\
&\times&\left\langle
N_f\frac{Q^2}{mV}\left(\frac{I_1}{m}\right)^{k_1+k_3}
\left(-I_2\right)^{k_2}
\left(-I_1\right)^{n_4}
\right\rangle_m
\end{eqnarray}
in the large volume limit, where zero modes contributions are neglected except for the first term.
According to the property Eq.~(\ref{eq:resultzero}) and the assumption
(\ref{eq:powerassumption2}), the RHS indeed vanishes 
in the $V\to\infty$ limit at small but non-zero $m$.
In the case with $(n_1, n_2, n_3)$=(odd, odd, even),
a similar analysis gives the same conclusion.

We conclude that, for a class of operators  
we have considered in this paper, the $U(1)_A$ breaking effects
are invisible in the thermodynamical limit.

%%%%%%%%%%%%%%%%%%%%%%%%%%%
\subsection{Possible phase diagrams including the strange quark}
%%%%%%%%%%%%%%%%%%%%%%%%%%%

Although we have so far discussed the $N_f=2$ case only,
it is interesting to consider possible phase diagrams
including the dynamical strange quark.
(In this subsection, let us denote the up and down quark mass
by $m_{ud}$ and the strange quark mass by $m_s$.)

Assuming that the $U(1)_A$ symmetry is still broken
above the critical temperature, a phase diagram
like the left panel of Fig.~\ref{fig:phase} is often shown
in the literature.
The quenched limit ($m_{ud}=m_s=\infty$) and
the $SU(3)$ symmetric chiral limit ($m_{ud}=m_s=0$) are
both expected to be in the first order transition regions,
while the physical point is located in the middle crossover region.
The critical curve around the $SU(3)$ limit has an end-point
at a finite value of $m_s$, from which 
a second order transition line (with $O(4)$ scaling \cite{Pisarski:1983ms}) 
is extended to the $N_f=2$ ($m_s=\infty$) limit.

Our new results may suggest a different diagram.
Since the $U(1)_A$ anomaly effects are invisible,
the chiral phase transition could be the first order.
Then, as shown in the right panel of Fig.~\ref{fig:phase},
one should have a critical value of the up and down quark mass
(let us denote this by $m_{ud}^{cr}$) from which the critical curve
may be extended to the finite $m_s$ region and
even connected to the curve 
around the first order transition region near $m_s=0$.

Since our study is limited to the $N_f=2$ case only,
the above scenario is just one example of 
many possible diagrams.
As pointed out in Refs.~\cite{Basile:2004wa, Basile:2005hw, Vicari:2007ma}
the second order transition is also possible.
But even in this case, its $U(2)_L\otimes U(2)_R/U(2)_V$ 
universality class is different from the
conventional $O(4)$ class.

Our simple analysis in the $N_f=2$ theory
thus suggests a richer structure in the QCD phase diagram.
It is particularly interesting for lattice QCD studies
to investigate the existence of $m_{ud}^{cr}$,
which may also be the boundary 
of the region where Eq.~(\ref{eq:resultzero}) holds\footnote{
Namely, the number of the exact zero modes
could be an order parameter.}.

\begin{figure*}[tb]
  %\centering
  \includegraphics[width=7.5cm]{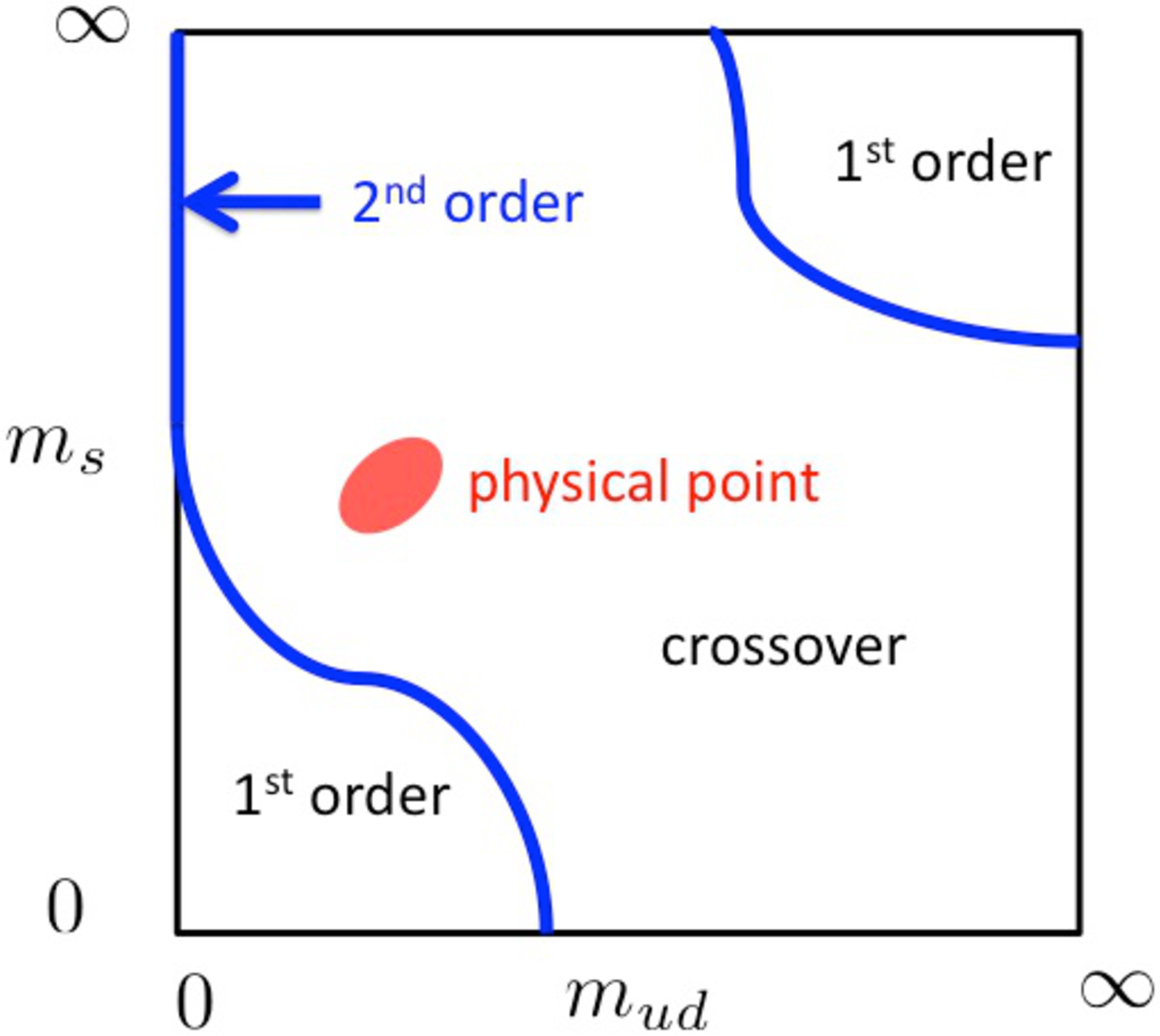}
  \includegraphics[width=7.5cm]{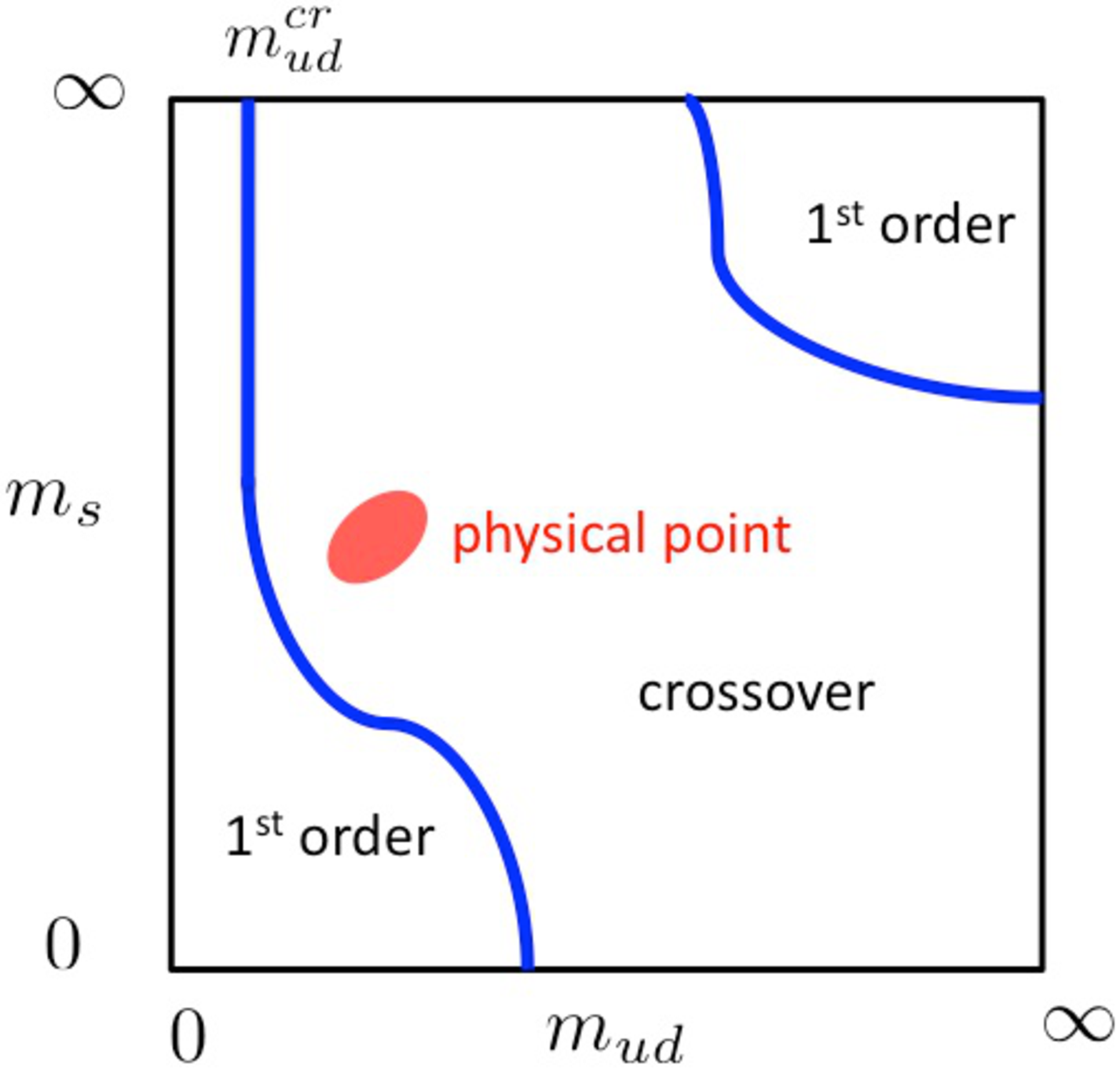}
\caption{
Possible phase diagrams including the strange quark. 
Left: A conventional diagram with the second order scenario 
in the $m_s\to \infty$ limit. 
Right: A possible diagram with the first order scenario
where the critical curve is smoothly connected to
the small $m_s$ region.
}
\label{fig:phase}
\end{figure*}
%%%%%%%%%%%%%%%%%%%%%%%%%%%%%%%%%%%%%%%%%%%%%%%%%
%%%%%%%%%%%%%%%%%%%%%%%%%%%%%%%%%%%%%%%%%%%%%%%%%
\section{Possible artifacts in lattice QCD}
\label{sec:syserr}
%%%%%%%%%%%%%%%%%%%%%%%%%%%%%%%%%%%%%%%%%%%%%%%%%
%%%%%%%%%%%%%%%%%%%%%%%%%%%%%%%%%%%%%%%%%%%%%%%%%

In the previous sections, we have investigated 
the symmetry restoration for $T>T_c$,
fully relying on the exact chiral $SU(2)_L\times SU(2)_R$ symmetry
and taking the thermodynamical limit $V\to \infty$.
However, numerical lattice QCD simulations
must be performed on a finite volume, sometimes employing fermion actions which explicitly break 
the chiral symmetry.
In this section, we would like to briefly address 
possible systematic effects of not having these two key properties.

First, we discuss the explicit breaking of 
the chiral $SU(2)_L\times SU(2)_R$ symmetry.
In order to characterize its violation, 
let us introduce a {\it mass} parameter $m_{\rm break}$.
Since $m_{\rm break}$ should disappear in the continuum limit,
it is natural to assume $m_{\rm break}\sim \Lambda_{\rm QCD}^2a$,
where $\Lambda_{\rm QCD}$ is the QCD scale.
At finite temperature $T$, we also have a possibility of
$m_{\rm break} \sim T^2 a$ or  $m_{\rm break} \sim \Lambda_{\rm QCD} T a$.
But it is unlikely that the lattice artifacts grow 
with $T$, since they are naively expected to be milder
in the weakly coupled region at higher temperature.
We also neglect the possibility of $m_{\rm break} \sim T_c^2 a$
since $T_c$ is not essentially different from $\Lambda_{\rm QCD}$.

If one employs improved Wilson-type actions
or staggered-type actions, it may be reduced to $O(a^2)$:
$m_{\rm break}\sim \Lambda_{\rm QCD}^3a^2$.
For the domain-wall fermion action,
as an approximation of the overlap fermion action,
the suppression of the discretization effects 
could be stronger.
In this case, the so-called residual mass, $m_{\rm res}$, 
is a good estimate for $m_{\rm break}$.
For this reason, we have introduced a rather abstract
parameter $m_{\rm break}$, to treat the conditions with 
different actions in a uniform manner.

Now the discussion is simple.
By losing the required exact chiral symmetry, 
every result in the previous sections
should be, in principle, modified 
by the effects of $m_{\rm break}$,
unless some special cancellation mechanism occurs.
The condition, 
$\langle \rho_0^A \rangle_m =0 $, could be an only exception,
being as a {\it definition} of the symmetry restoration.

Namely, instead of Eqs.~(\ref{eq:resultrho}), 
we should have
\begin{eqnarray}
\langle \rho^A(\lambda)\rangle_m 
&=& \alpha m_{\rm break} \Lambda_{\rm QCD}\lambda + \beta m_{\rm break} \lambda^2 + 
\left(\langle \rho_3^A\rangle_0 +\gamma m_{\rm break}/\Lambda_{\rm QCD}\right)
\frac{\lambda^3}{3!}+\cdots, 
\end{eqnarray} 
where $\alpha, \beta, \gamma, \cdots$ are unknown 
dimensionless $O(1)$ coefficients. 
Similarly,  $\langle N_{R+L}\rangle_m/V$, $\langle Q(A)^2\rangle_m/V$, $\chi^{\pi-\eta}$ and so on should not be zero but $O(m_{\rm break})$.
Note that there is no reason for the $U(1)_A$ symmetry to get restored as long as the chiral symmetry is explicitly broken by the quark action at finite lattice spacings.

Next, let us discuss the finite volume effects.
Even in the continuum limit, it is possible to
obtain different results from what we have shown in this paper.
It is good to remember that Eq.~(\ref{eq:LeeHatsuda})
is an example of the $U(1)_A$ breaking in the continuum theory,
as a finite volume effect.
The recovery of the $U(1)_A$ symmetry is not as strong as 
the other symmetries, under which the Lagrangian is strictly constrained,
but rather a consequence from QCD dynamics, which is manifest only in the thermodynamical limit.

Above the critical temperature, the long-range physics of the system 
would be characterized by the correlation length $\xi$,
or the inverse of some screening mass, which diverges at $T_c$ if the phase transition is of second order while
remaining finite for the first order transition.

It is then natural to assume that a lattice QCD simulation 
has  finite volume effects as 
functions of $\xi/V^{1/4}$.
It is also important to note that, unlike the
truly local quantity, whose
finite volume effects are exponential $\sim \exp(-V^{1/4}/\xi)$,
the susceptibilities considered in this paper 
are volume averaged quantities, so that any finite volume effect
is expected to be a power function of $\xi/V^{1/4}$.
A careful estimate for the thermodynamical limit
is thus required, in particular, for temperature near $T_c$, where $\xi$ could become larger.

%%%%%%%%%%%%%%%%%%%%%%%%%%%%%%%%%%%%%%%%%%%%%%%%%
%%%%%%%%%%%%%%%%%%%%%%%%%%%%%%%%%%%%%%%%%%%%%%%%%
\section{Eigenvalue density with fractional power}
\label{sec:fractional}
%%%%%%%%%%%%%%%%%%%%%%%%%%%%%%%%%%%%%%%%%%%%%%%%%
%%%%%%%%%%%%%%%%%%%%%%%%%%%%%%%%%%%%%%%%%%%%%%%%%

So far, we have assumed that $\rho^A(\lambda)$ 
is analytic around $\lambda=0$, and have used 
the expansion in Eq.~(\ref{eq:EV_exp}).
In this section, let us extend our analysis to 
a nonanalytic case where
\beqa
\rho^A(\lambda) = c^A \lambda^\gamma,
\label{eq:eigen_frac}
\eeqa 
with a fractional power $\gamma$ for $\lambda < \epsilon$, where $c^A$ is an $A$-dependent constant.
Since $\lim_{m\rightarrow 0}\langle\rho^A(0)\rangle_m=0$ in the $SU(2)_L\times SU(2)_R$ symmetric phase, $\gamma$ should be positive as long as $\langle c^A\rangle_m = O(1)$.
It is still true in this case
that only the vicinity of $\lambda=0$ contributes to the WTI's.
We thus can neglect additional terms with
higher order fractional powers, even if they exist
in the bulk region of $\lambda\ge \epsilon$.

In this case, $I_1$, $I_2$, and $I_3$ are expressed as
\begin{eqnarray}
I_1 &=&  
c^A \left[m^\gamma \left( d_1 + O(m^2)\right) + m\left( e_1+O(m^2)\right) \right],
\\
\frac{I_1}{m}+I_2 &=& 2mI_3
= c^A \left[
m^{\gamma-1}\left( d_2  + O(m^2)\right)
+ m^2 \left(e_2 + O(m^2)\right]
\right) ,
\end{eqnarray}
where $d_i$'s and $e_i$'s are given by
\begin{eqnarray}
d_1 &=& 
\pi\sec\left(\frac{\gamma \pi}{2}\right),\quad
d_2 =
(1-\gamma)\pi\sec\left(\frac{\gamma \pi}{2}\right),
\end{eqnarray}
and 
\begin{eqnarray}
e_1 &=&\int_\epsilon^{\Lambda_R}d\lambda\,\rho^A(\lambda)\frac{2g_0(\lambda^2)}{Z_m^2\lambda^2+m^2}+
\epsilon^{\gamma-1}\frac{
\Gamma \left(\frac{\gamma-1}{2}\right)
-\frac{\epsilon^2}{\Lambda_R^2}\frac{\Gamma \left(\frac{\gamma +1}{2}\right)^2}
{\Gamma\left(\frac{\gamma +3}{2}\right)}}
{\Gamma \left(\frac{\gamma+1}{2}\right) },
\\
e_2 &=&
\int_\epsilon^{\Lambda_R}d\lambda\,\rho^A(\lambda)\frac{4 g_0^2(\lambda^2)}{(Z_m^2\lambda^2+m^2)^2}
+
\epsilon^{\gamma-3}
\frac{ (\gamma+3) \Gamma \left(\frac{\gamma-1}{2}\right)}
{2(\gamma-3) \Gamma \left(\frac{\gamma+5}{2}\right)} \nn \\
&&\times
\left[(\gamma+1)(\gamma-1) -2\frac{\epsilon^2}{\Lambda_R^2}(\gamma+1)(\gamma-3)
+\frac{\epsilon^4}{\Lambda_R^4}(\gamma-1)(\gamma-3)
\right] , 
\end{eqnarray}
with the UV cut-off $\Lambda_R$ and the IR cut-off $\epsilon$.
Note that the $d_i$'s and $e_i$'s are all finite.

%%%%%%%%%%%%%%%%%%%%%%%%%%%%%%%%%%%%%%%%%%%%%%%%%
\subsection{$0<\gamma<1$}
%%%%%%%%%%%%%%%%%%%%%%%%%%%%%%%%%%%%%%%%%%%%%%%%%

We first consider the case with $\gamma < 1$.
With the above expressions for the $I_i$'s, let us reexamine the WTI's
given in the previous sections.

For  $\langle S_0^N\rangle_m /V^N$ with an arbitrary $N$,
the WT identity requires
\beqa
\lim_{m\rightarrow 0}\lim_{V\rightarrow\infty}\left\langle \left\{\frac{N_{R+L}^A}{mV} + I_1\right\}^N\right\rangle_m = 0,
\eeqa
where the positivity of each term implies
\beqa
\label{eq:NRLfractional}
\lim_{V\rightarrow\infty}\frac{\langle N_{R+L}^A\rangle_m}{V} = 0,
\eeqa
at small $m$, and
\beqa
\lim_{m\rightarrow 0} m^{N\gamma} d_1^N \langle (c^A)^N \rangle_m = 0,
\eeqa
which is automatically satisfied for positive $\gamma$.

At $N=2$, we have
\beqa
\left\langle \frac{I_1}{m} + I_2 -\frac{N_f Q(A)^2}{m^2 V} \right\rangle_ m
&=& d_2m^{\gamma-1}\langle c^A\rangle_m -\frac{N_f\langle  Q(A)^2\rangle_m }{m^2 V}\rightarrow 0.
\label{eqn:N2cA}
\eeqa
Taking into account a fact that both $c^A$ and $Q^2$ are mass independent
and their expectation value should be written as an even power of mass,
both terms in Eq.~(\ref{eqn:N2cA}) should vanish separately,
which
leads to
\beqa
\langle c^A\rangle_m = O(m^2), \quad
\frac{\langle Q(A)^2\rangle_m}{V} = O(m^4).
\eeqa

At $N=3$, it is not difficult to see that 
all the non-trivial conditions are automatically satisfied
with the above constraints.

At $N=4$ there remains one non-trivial WT identity:
\beqa
\langle 6mI_3(I_2-I_1/m)\rangle_m + \frac{6 N_f\langle  I_1 Q(A)^2\rangle_m }{m^3 V}
-\frac{N_f^2 \langle Q(A)^4\rangle_m}{m^4 V^2} \rightarrow 0 .
\eeqa
Since the first two terms  vanish in the chiral limit,  
we obtain a new constraint that
\beqa
\lim_{V\rightarrow \infty}\frac{\langle Q(A)^2 \rangle_m}{V} = O(m^6) .
\eeqa

Let us finally consider the WT identity from 
${\cal O}_a^{N=4k} ={\cal O}_{0,1,(4k-1),0}$ as before.
In a way very similar to that in Section~\ref{subsec:special},
we can show
\beqa
\label{eq:Qfractional}
\lim_{V\rightarrow\infty}
\frac{\langle Q(A)^2\rangle_m}{  V} &=& 0, \\
\langle c^A \rangle_m &=& 0,
\eeqa
even at non-zero $m$. 
This means that the Dirac eigenvalue density with a 
fractional power, Eq.~(\ref{eq:eigen_frac}), is incompatible 
with the $SU(2)_L\times SU(2)_R$ chiral symmetry restoration 
for $0< \gamma < 1$.

%%%%%%%%%%%%%%%%%%%%%%%%%%%%%%%%%%%%%%%%%%%%%%%%%
\subsection{$1< \gamma < 2$}
%%%%%%%%%%%%%%%%%%%%%%%%%%%%%%%%%%%%%%%%%%%%%%%%%

Next, let us consider $1< \gamma < 2$.
Our strategy is the same as in the previous subsection,
except that the leading term is not $O(m^\gamma)$,
 but rather is $O(m)$ in $I_1$.

Up to $N=4$, one can easily confirm that most of the 
conditions are automatically satisfied for $1<\gamma<2$,
keeping the constraints on the zero-mode contribution
Eqs.~(\ref{eq:NRLfractional}) and (\ref{eq:Qfractional})
unchanged.
The only non-trivial WT identity appears at $N=3$:
\begin{eqnarray}
\lim_{V\to \infty}\frac{\langle {\cal O}_{1110}\rangle_m}{V}
&=& 2N_f \langle I_3\rangle_m
= 2N_f d_3 m^{\gamma-2}\langle c^A\rangle_m +O(m) = 0,
\end{eqnarray}
which leads to a constraint
\begin{equation}
\langle c^A\rangle_m = O(m^2).
\end{equation}
Namely, the fractional power $\gamma < 2$
cannot survive in the chiral limit.

%%%%%%%%%%%%%%%%%%%%%%%%%%%%%%%%%%%%%%%%%%%%%%%%%
\subsection{$2< \gamma < 3$}
%%%%%%%%%%%%%%%%%%%%%%%%%%%%%%%%%%%%%%%%%%%%%%%%%

In this case, all the conditions from WTI's 
are  automatically satisfied up to $N=6$ as long as Eqs.~(\ref{eq:NRLfractional}) and (\ref{eq:Qfractional})
are satisfied.
We thus have no constraint on $\langle c^A \rangle$.

However, it is important to note that
excluding $\gamma \le 2$ in the chiral limit is enough
to achieve all the $U(1)_A$ symmetric identities
in Section \ref{sec:singlet}.
As discussed in subsection \ref{subsec:U(1)generalN},
the zero-mode's contribution plays a more
important role than bulk contributions from non-zero modes.

%%%%%%%%%%%%%%%%%%%%%%%%%%%%%%%%%%%%%%%%%%%%%%%%%
%%%%%%%%%%%%%%%%%%%%%%%%%%%%%%%%%%%%%%%%%%%%%%%%%
\section{Summary and discussion}
\label{sec:conclusion}
%%%%%%%%%%%%%%%%%%%%%%%%%%%%%%%%%%%%%%%%%%%%%%%%%
%%%%%%%%%%%%%%%%%%%%%%%%%%%%%%%%%%%%%%%%%%%%%%%%%

In this paper, we have investigated 
the eigenvalue density $\rho^A(\lambda)$ of the Dirac operator
in the chiral $SU(2)_L\times SU(2)_R$ symmetric phase
at finite temperature.
In order to avoid possible ultra-violet
divergences, we have worked analytically on a lattice, 
employing the overlap Dirac operator, which ensures 
the exact chiral symmetry at finite lattice spacings.

From the various WT identities 
of the scalar and pseudoscalar operators, 
we have shown that a behavior such as
$\langle\rho^A(\lambda)\rangle_m \propto \lambda^\gamma$ for small $\lambda$
cannot survive in the chiral limit for $\gamma \le 2$.
If $\langle\rho^A(\lambda)\rangle_m$ is analytical
around $\lambda=0$, this means that 
it should start with a cubic term, as is the case
with the free quark theory.
Moreover, we have found a strong suppression
on the zero-mode's contributions in the thermodynamical limit.
As shown in Eq.~(\ref{eq:resultzero}),
they disappear even with small but finite $m$.
It is worth mentioning that the use of the overlap fermion is crucial for obtaining the results in this paper since only this fermion formulation can preserve the exact chiral symmetry with non-perturbative cut-off, which makes our arguments more rigorous.

The obtained constraints on the Dirac spectrum
are strong enough for all of the $U(1)_A$ 
breaking effects among correlation functions of scalar and pseudo scalar operators  considered in this work 
to vanish in the limits of $V\to \infty$ and then $m\to 0$.
Namely, there is no remnant of the $U(1)_A$ anomaly 
above the critical temperature
at least in these correlation functions.

This does not contradict with the apparently opposite results 
about the $U(1)_A$ restoration in the previous works.
As we have shown in Section~\ref{sec:Intro},
their $U(1)_A$ breaking parts cannot survive
in the thermodynamical limit ($V\to \infty$),
but they could be finite on a finite box,
which may be a part of the difficulties of numerical 
lattice QCD simulations.

We only use a part of the chiral Ward-Takahshi identities to derive the constraints in this paper, which are therefore necessary conditions to be fulfilled if the chiral symmetry is restored. Our results strongly rely on analyticity in $m^2$ for $m$-independent observables and its consequence eq.~(\ref{eq:m-dep}). If our results are shown to be incorrect by some numerical simulations, these assumptions must also be violated in the simulations.  

One of the most important consequence of our study is that, 
since the $U(1)_A$ anomaly effect disappears in scalar and pseudo scalar sectors at $T_c$, the chiral phase transition for 2 flavor QCD is likely to be of first order\cite{Pisarski:1983ms} or of second order in the $U(2)_L\otimes U(2)_R/U(2)_V$  universality class\cite{Basile:2004wa,Basile:2005hw,Vicari:2007ma},
contrary to the expectation that the chiral phase transition of 2 flavor QCD belongs to the $O(4)$ universality class.

\section*{Acknowledgements}
We thank members of JLQCD collaborations, in particular, Drs. T.~Onogi and G.~Cossu, for discussions and useful comments, and Drs. S.~Yamaguchi, and K.~Kanaya for useful discussions.
We would especially like to thank Dr. E. Vicari for pointing out  the existence of the $U(2)_L\otimes U(2)_R/U(2)_V$  universality class.
We also thank the Galileo Galilei Institute for Theoretical Physics for its kind hospitality during completion of this paper while attending the workshop  ``New Frontiers Lattice Gauge Theory'' .
This work is supported in part by  the Grant-in-Aid of the Japanese Ministry of Education (No. 22540265), 
the Grant-in-Aid for Scientific Research on Innovative Areas 
(No. 2004: 20105001, 20105003, 23105710,  23105701) and  SPIRE (Strategic Program for Innovative Research).

\appendix
\section{Useful formulas}

\subsection{Quark contractions for the (pseudo) scalar operator}
Here we give a contraction formula for the (pseudo) scalar operators when we
integrate out the fermion fields:
\begin{eqnarray}
\langle S_0 \rangle_F &=& -N_f\tr\, \tilde S_A, \qquad 
\langle P_0 \rangle_F = -iN_f\tr\, \gamma_5 \tilde S_A, 
\nn \\ 
\langle S_a^2 \rangle_F &=& -N_f\tr\, \tilde S_A^2,
\qquad 
\langle P_a^2 \rangle_F = N_f\tr\, \left(\gamma_5 \tilde S_A\right)^2,\nn
\\
\langle S_0^2 \rangle_F &=& -N_f\tr\, \tilde S_A^2
+\left(N_f\tr\, \tilde S_A\right)^2,\qquad 
\langle P_0^2 \rangle_F = N_f\tr\, \left(\gamma_5 \tilde S_A\right)^2
-\left(N_f\tr\, \gamma_5 \tilde S_A\right)^2, \nn
\\
\langle S_0P_0 \rangle_F &=& -iN_f\tr\, \gamma_5\tilde S_A^2
+iN_f^2\tr\, \tilde S_A \tr\, \gamma_5 \tilde S_A,\qquad 
\langle S_aP_a \rangle_F = -iN_f\tr\, \gamma_5\tilde S_A^2 ,
\end{eqnarray}
where $\tilde S_A(x,y) \equiv F(D) S_A(x,y)$.

\subsection{Trace of fermion propagators}
\label{sec:trace}
Here we give useful formulas for the trace of fermion propagators 
in a form of the eigenvalue decomposition.

Let us first define $S_A^n$ as
\beqa
\tilde S_A^n &\equiv &\int d^4x_1 d^4x_2\cdots d^4x_n\, \tilde S_A(x_1,x_2) \tilde S_A(x_2,x_3)\cdots \tilde S_A(x_n,x_1)\nn \\
& =&\int \prod_{i=1}^n  d^4x_i\, \tilde S_A(x_i, x_{i+1}), \quad (x_{n+1}=x_1) .
\eeqa

Inserting the eigenvalue decomposition for the fermion propagator 
Eq.~(\ref{eq:spectrumS}), we obtain
\beqa
\frac{1}{V}\tr\, \tilde S_A &=& 
 \frac{N_{R+L}^A}{mV} + I_1, \quad
 \frac{1}{V}\tr\, \tilde S_A^2 = 
 \frac{N_{R+L}^A}{m^2 V} + I_2, \\
\frac{1}{V}\tr\, (\gamma_5 \tilde S_A)^2 &=& 
 \frac{N_{R+L}^A}{m^2 V} + \frac{I_1 }{m},\quad
\frac{1}{V}\tr\, \gamma_5 \tilde S_A\gamma_5 \tilde S_A^2 =
\frac{N_{R+L}^A}{m^3 V} + I_3, \\
\frac{1}{\sqrt{V}}\tr\,  \gamma_5 \tilde S_A &=& \frac{Q(A)}{m\sqrt{V}},  \quad
\frac{1}{\sqrt{V}}\tr\, \gamma_5 \tilde S_A^2 = \frac{Q(A)}{m^2\sqrt{V}} ,
\eeqa
where $I_i$ ($i=1,2,3$) are expressed in terms of the eigenvalue density
in the large volume limit as 
\begin{eqnarray}
I_{2k-1}&=&
m\int_{0}^{\Lambda_R}d\lambda\rho^A(\lambda)
\frac{2g_0^k(\lambda^2)}{\left(Z_m^2\lambda^2+m^2\right)^k},
\\
g_0(\lambda^2)&=&1-\frac{\lambda^2}{\Lambda_R^2},\quad
Z_m^2=1-\frac{m^2}{\Lambda_R^2},
\\
I_2&=&-\frac{I_1}{m}+2mI_3.
\end{eqnarray}

\subsection{Various integrals of eigenvalue density}
\label{sec:integral}
The above $I_i$'s are evaluated by expanding the eigenvalue density
as Eq.~(\ref{eq:EV_exp}).
In evaluating $I_{2k-1}$ it may be better to rewrite $I_{2k-1}=I_{2k-1}^\epsilon + O(m)$ where
\begin{eqnarray}
I_{2k-1}^\epsilon&=&
\frac{m_R}{Z_m^{2k-1}}\int_{0}^{\epsilon}d\lambda\rho^A(\lambda)
\frac{2g_0^k(\lambda^2)}{\left(\lambda^2+m_R^2\right)^k}
\end{eqnarray}
by using $m_R=m/Z_m$.
The expansion is given by
\begin{eqnarray}
I_{2k-1}^\epsilon&=&
\sum_{n=0}^{\infty}\rho_n^AI_{2k-1}^{(n)}
\end{eqnarray}
with the expansion coefficient
\begin{eqnarray}
I_{2k-1}^{(n)}=
\frac{1}{Z_m^{2k-1}}\int_{0}^{\epsilon}d\lambda\frac{\lambda^n}{n!}
\frac{2m_Rg_0^k(\lambda^2)}{\left(\lambda^2+m_R^2\right)^k}.
\end{eqnarray}
At $n > 2k-1$ the leading term in $m$ is given by
\begin{eqnarray}
I_{2k-1}^{(n){\rm leading}}&=&
2m_R\int_0^\epsilon\frac{\lambda^{n-2k}}{n!} \left(1-\frac{\lambda^2}{\Lambda_R^2}\right)^k
\end{eqnarray}

The explicit form of the coefficient is given as follows for a few $k$
and $n$
\begin{eqnarray}
I_1^{(0)}&=&
\frac{2}{Z_m} \left[\left(1+\frac{m_R^2}{\Lambda_R^2}\right)
 \tan^{-1}\left(\frac{\epsilon}{m_R}\right)-\frac{m_R\epsilon}{\Lambda_R^2}\right],
\\
I_1^{(1)}&=&
\frac{m_R}{Z_m} \left[\left(1+\frac{m_R^2}{\Lambda_R^2}\right)
 \log \left(\frac{\epsilon^2}{m_R^2}+1\right)-\frac{\epsilon^2}{\Lambda_R^2}\right],
\\
I_1^{(2)}&=&
\frac{m_R}{Z_m}
\left[\epsilon\left(1+\frac{3m_R^2-\epsilon^2}{3\Lambda_R^2} \right)-m_R \left(1+\frac{m_R^2}{\Lambda_R^2}\right)
 \tan^{-1}\left(\frac{\epsilon }{m_R}\right)\right],
\\
I_3^{(0)}&=&
\frac{1}{Z_m^3m_R^2}\left[
\left(1-3 \frac{m_R^2}{\Lambda_R^2}\right)
 \left(1+\frac{m_R^2}{\Lambda_R^2}\right)
\tan^{-1}\left(\frac{\epsilon}{m_R}\right)
+\frac{m_R\epsilon}{m_R^2+\epsilon^2} \left(1+ \frac{m_R^2}{\Lambda_R^2}\right)^2
+\frac{2m_R^3\epsilon}{\Lambda_R^4}
\right], \nn
\\
\\
I_3^{(1)}&=&
\frac{m_R}{Z_m^3}
\left[
\frac{\epsilon^2}{ m_R^2(\epsilon^2+m_R^2)}\left(1+\frac{m_R^2}{\Lambda_R^2}\right)^2
+\frac{\epsilon^2}{\Lambda_R^4}
+\frac{2}{\Lambda_R^2}\left(1+\frac{m_R^2}{\Lambda_R^2}\right)
 \log \left(\frac{m_R^2}{\epsilon^2+m_R^2}\right)\right],
\\
I_3^{(2)}&=&
\frac{1}{2Z_m^3}
\left[
\left(1+\frac{m_R^2}{\Lambda_R^2}\right)
\left(1+5\frac{m_R^2}{\Lambda_R^2}\right)
\tan^{-1}\left(\frac{\epsilon}{m_R}\right)+\frac{2m_R\epsilon^3}{3\Lambda_R^4}\right. \nn \\
& &\left. - \frac{m_R\epsilon}{\Lambda_R^2}\left(1+\frac{m_R^2}{\Lambda_R^2}\right)\left(4+\frac{\Lambda_R^2+m_R^2}{\epsilon^2+m_R^2}\right)\right],
\\
I_3^{(3)}&=&
\frac{m_R}{6Z_m^3} \left[
\left(1+\frac{m_R^2}{\Lambda_R^2}\right)
\left(1+3\frac{m_R^2}{\Lambda_R^2}\right)
 \log \left(\frac{\epsilon^2}{m_R^2}+1\right)
 +\frac{\epsilon^4}{2\Lambda_R^4}\right. \nn \\
&&\left. - \frac{\epsilon^2}{\Lambda_R^2}\left(1+\frac{m_R^2}{\Lambda_R^2}\right)\left(2+\frac{\Lambda_R^2+m_R^2}{\epsilon^2+m_R^2}\right)\right] .
\end{eqnarray}
According to the equality the coefficient for $I_2$ is given by
\begin{eqnarray}
I_2^{(0)}&=&
\frac{2\epsilon\left(\frac{(\Lambda_R^2+m_R^2)^2}{\epsilon^2+m_R^2} +2m_R^2+\Lambda_R^2\right)-6 m_R \left(\Lambda_R^2+m_R^2\right) \tan^{-1}
\left(\frac{\epsilon}{m_R}\right)}{\Lambda_R^4 Z_m^2},
\\
I_2^{(1)}&=&
\frac{\epsilon^2\left(\frac{2(\Lambda_R^2+m_R^2)^2}{\epsilon^2+m_R^2} +2m_R^2+\Lambda_R^2\right)-(\Lambda_R^2+4m_R^2)
 \left(\Lambda_R^2+m_R^2\right) \log \left(\frac{\epsilon^2}{m_R^2}+1\right)
}{\Lambda_R^4 Z_m^2},
\\
I_2^{(2)}&=&\frac{1}{3 \Lambda_R^4 Z_m^2}
\left[ 3m_R (\Lambda_R^2+m_R^2)(2\Lambda_R^2+5m_R^2)
 \tan^{-1}\left(\frac{\epsilon }{m_R}\right)\right. \nn\\
 &&\left.
-\epsilon  \left(3( \Lambda_R^2+m_R^2)\left\{\Lambda_R^2+4m_R^2 +m_R^2\frac{\Lambda_R^2+m_R^2}{\epsilon^2+m_R^2}\right\}-\epsilon^2(\Lambda_R^2+2m_R^2)\right)\right] .
\end{eqnarray}

\subsection{Integrals of eigenvalue density with fractional power}
\label{sec:integral_frac}
If we consider the fractional power for the eigenvalue density with
$\gamma>0$ the eigenvalue integral
\begin{eqnarray}
I_{2k-1}^{(\gamma)}&=&
\frac{m_R}{Z_m^{2k-1}}\int_{0}^{\epsilon}d\lambda\, \lambda^\gamma
\frac{2g_0^k(\lambda^2)}{\left(\lambda^2+m_R^2\right)^k}
\end{eqnarray}
is given in terms of the hyper geometric function as
\begin{eqnarray}
I_1^{(\gamma)}&=&
   \frac{2 \epsilon^{\gamma +1} \left((\gamma +3)
 _2F_1\left(1,\frac{\gamma+1}{2};\frac{\gamma +3}{2};-\frac{\epsilon^2}{m_R^2}
\right)
-\displaystyle\frac{\epsilon^2}{\Lambda_R^2}(\gamma +1) _2F_1\left(1,\frac{\gamma +3}{2};\frac{\gamma +5}{2};
-\frac{\epsilon^2}{m_R^2}\right)\right)}{(\gamma +1) (\gamma +3) Z_m m_R}, \nn \\
\\
I_3^{(\gamma)}&=&
\frac{\epsilon^{\gamma+1}}{m_R^3 Z_m^3}\left[
\frac{m_R^2}{\epsilon^2+m_R^2}g_0(\epsilon^2)^2
-\frac{\gamma-1}{\gamma+1}\,  {}_2F_1\left(1,\frac{\gamma +1}{2};\frac{\gamma +3}{2};
-\frac{\epsilon^2}{m_R^2}\right)\right. \nn \\
&&\left.
-\frac{2\epsilon^2}{\Lambda_R^2}\frac{\gamma+1}{\gamma+3}\,  {}_2F_1\left(1,\frac{\gamma +3}{2};\frac{\gamma +5}{2};
-\frac{\epsilon^2}{m_R^2}\right)
-\frac{\epsilon^4}{\Lambda_R^4}\frac{\gamma+3}{\gamma+5} \, {}_2F_1\left(1,\frac{\gamma +5}{2};\frac{\gamma +7}{2};
-\frac{\epsilon^2}{m_R^2}\right)
\right] , \nn \\
\end{eqnarray}
where ${}_2F_1$ is the Gaussian hyper-geometric function given by
\begin{eqnarray}
{}_2F_1(\alpha,\beta,\gamma;z)
=\frac{\Gamma(\gamma)}{\Gamma(\alpha)\Gamma(\beta)}
\sum_{n=1}^{\infty}\frac{\Gamma(\alpha+n)\Gamma(\beta+n)}{\Gamma(\gamma+n)}
\frac{z^n}{n!}.
\end{eqnarray}

Performing an expansion for $m_R/\epsilon \ll1$ we have
\begin{eqnarray}
I_1^{(\gamma)}&=&
\frac{m_R}{\epsilon}
   \frac{\epsilon^{\gamma } \left(\Gamma \left(\frac{\gamma
    -1}{2}\right)-\displaystyle\frac{\epsilon^2}{\Lambda_R^2}\frac{\Gamma \left(\frac{\gamma +1}{2}\right)^2}{\Gamma
    \left(\frac{\gamma +3}{2}\right)}\right)}
{Z_m \Gamma \left(\frac{\gamma +1}{2}\right)}
+O\left(\left(\frac{m_R}{\epsilon}\right)^3\right)
\nn\\&&
+\left(\frac{m_R}{\epsilon}\right)^{\gamma }
\Biggl[
\frac{2 \epsilon^{\gamma } \Gamma \left(\frac{1}{2}-\frac{\gamma }{2}\right)
 \Gamma \left(\frac{\gamma }{2}+\frac{3}{2}\right)}{(1+\gamma)Z_m}
+O\left(\left(\frac{m_R}{\epsilon}\right)^2\right)
\Biggr],
\\
I_3^{(\gamma)}&=&
\frac{ (\gamma +3) \epsilon^{\gamma -2} \Gamma \left(\frac{\gamma
    -1}{2}\right) \left(\displaystyle\frac{m_R}{\epsilon}\right)}{4(\gamma -3) Z_m^3 \Gamma \left(\frac{\gamma
    +5}{2}\right)}\left[(\gamma+1)(\gamma-1)-\frac{2\epsilon^2}{\Lambda_R^2}(\gamma+1)(\gamma-3)+\frac{\epsilon^4}{\Lambda_R^4}(\gamma-1)(\gamma-3)
    \right]\nn \\
    &+&O\left(\left(\frac{m_R}{\epsilon}\right)^2\right)
+\left(\frac{m_R}{\epsilon}\right)^{\gamma-2 }
\Biggl[
\frac{
    (1-\gamma ) \epsilon^{\gamma -2} \Gamma
    \left(\frac{1-\gamma}{2}\right) \Gamma \left(\frac{1+\gamma
    }{2}\right)}{2 Z_m^3
  }
+O\left(\left(\frac{m_R}{\epsilon}\right)^2\right)
\Biggr].
\end{eqnarray}

\subsection{General correlation functions in large volume limit}
\label{sec:volume}
Here we consider the leading volume scaling of 
the general correlation functions made of $S_a$'s and $P_a$'s.

At the given order $N=2(k_1+k_2+k_3)+n_4$,
there are two types of parity-even and chiral symmetric operators,
\beqa
{\cal O}_1^N &=& {\cal O}_{2k_1,2k_2,2k_3, n_4}, \quad
{\cal O}_2^N = {\cal O}_{2k_1+1,2k_2+1,2k_3+1, n_4-3} .
\eeqa
For ${\cal O}_1^N$, the integration over the fermion fields 
in the large volume limit is given by 
\beqa
\langle {\cal O}_1^N \rangle_F &\simeq& \langle P_a^{2k_1} \rangle_F   \langle S_a^{2k_2} \rangle_F   \langle P_0^{2k_3} \rangle_F  \langle S_0 \rangle_F^{\, n_4}  \nn \\
&\sim&   \langle P_a^{2} \rangle_F^{\, k_1}  \langle S_a^{2} \rangle_F^{\, k_2}   \langle P_0^{2} \rangle_F^{k_3}  \langle S_0 \rangle_F^{\, n_4} ,
\eeqa
where each $\langle {\cal O}\rangle_F$ gives an $O(V)$ contribution.
An overall  constant coming from combinatorial factors is omitted here and hereafter. Therefore the leading contribution is $O(V^{k_1+k_2+k_3+n_4})$ 
for ${\cal O}_1^N$.

Similarly we have 
\beqa
\langle {\cal O}_2^N \rangle_F &\sim&  \langle P_aS_a\rangle_F 
\langle P_a^{2} \rangle_F^{\, k_1}  \langle S_a^{2} \rangle_F^{\, k_2}   \langle P_0^{2k_3+1} \rangle_F  \langle S_0 \rangle_F^{\, n_4-3} ,
\eeqa
where 
\beqa
 \langle P_0^{2k_3+1} \rangle_F  &\sim & \langle P_0\rangle_F   \langle P_0^{2k_3} \rangle_F 
 \sim  \langle P_0\rangle_F   \langle P_0^2 \rangle_F^{k_3} .
\eeqa
Since $ \langle P_aS_a\rangle_F $ and $\langle P_0\rangle_F$ are proportional to $Q(A)$ and is therefore
$O(\sqrt{V})$, 
the leading volume dependence of ${\cal O}_2^N$ is $O(V^{k_1+k_2+k_3+n_4-2})$.

\section{Structure of Ward-Takahashi identities for scalar and pseudo-scalar operators}
\label{sec:general}
In this appendix, we summarize the general structures of 
WT identities among the scalar and pseudo-scalar operators,
${\cal O}_{n_1,n_2,n_3,n_4} = P_a^{n_1} S_a^{n_2} P_0^{n_3} S_0^{n_4}$.

In this work, we study the relation between three operator sets below,
\beqa
{\cal O}_a^{(N)} &\equiv& \{ {\cal O}_{n_1,n_2,n_3,n_4} \vert \, n_1+n_2={\rm odd},\ n_1+n_3={\rm odd},\
\sum_i n_i =N\}, \\
{\cal O}_0^{(N)} &\equiv& \{ {\cal O}_{n_1,n_2,n_3,n_4} \vert \, n_1+n_2={\rm even},\ n_1+n_3={\rm odd},\
\sum_i n_i =N\},  \\
{\cal O}^{(N)} &\equiv& \{ {\cal O}_{n_1,n_2,n_3,n_4} \vert \, n_1+n_2={\rm even},\ n_1+n_3={\rm even},\
\sum_i n_i =N\}. 
\eeqa
Note that the only ${\cal O}^{(N)}$ can have a non-zero 
expectation value in QCD with 2 degenerate quarks.

Since 
\beqa
\delta_a {\cal O}_a^{(N)} &\equiv& \{ \delta_a {\cal O} \vert {\cal O}  \in {\cal O}_a^{(N)}\}
\in {\cal O}^{(N)},   \qquad
\delta_0 {\cal O}_0^{(N)} \equiv \{ \delta_0 {\cal O} \vert {\cal O}  \in {\cal O}_0^{(N)}\} \in {\cal O}^{(N)},
\eeqa
our goal of this paper is to understand constraints
from the $SU(2)_L\times SU(2)_R$ symmetry restoration, 
$\langle \delta_a {\cal O}_a^{(N)}\rangle =0$, and to examine whether 
$\delta_0 {\cal O}_0^{(N)}$ can have 
a non-zero expectation value, 
which means that the $U(1)_A$ is still broken.
For simplicity, hereafter we denote $n=(n_1n_2n_3n_4)$ 
instead of ${\cal O}_{n_1,n_2,n_3,n_4}$ to represent an operator.

\subsection{WTI's at odd $N$}
As shown in Section~\ref{sec:singlet},
we can show that $\langle \delta_0 {\cal O}_0^{(N)}\rangle=0 $
if $\langle \delta_a {\cal O}_a^{(N)}\rangle=0 $ when $N$ is odd.
This follows from the fact that 
$\vert  {\cal O}_a^{(N)} \vert=\vert
{\cal O}_0^{(N)}  \vert=\vert{\cal O}^{(N)}\vert$, where 
 $\vert {\cal O}\vert$ means a number of independent operators in ${\cal O}$.
Here we give a proof for this equality for general odd $N=2k+1$.

At $k=0$, we have only one operator for each set:
$n_A=(1000)$ for ${\cal O}_a^{(N)}$, $n_B=(0010)$ for ${\cal O}_0^{(N)}$, 
and $n_C=(0001)$ for ${\cal O}^{(N)}$.
Thus, $\vert  {\cal O}_a^{(N)} \vert=\vert
{\cal O}_0^{(N)}  \vert=\vert{\cal O}^{(N)}\vert=1$ for $k=0$.

At $k=1$, we can create the operators by adding to the above 
$n_X(X=A,B,C)$ a pair of the {\it same} operators, namely, 
adding 2 to one element of $n_X$. 
Since each $n_X$ has 4 elements, we have 4 operators for each set.
We should, however, note that there is one additional type of 
operators for each set: $\bar n_A=(0111) \in {\cal O}_a^{(N)}$, 
$\bar n_B =(1101) \in {\cal O}_0^{(N)}$, and  
$\bar n_C=(1110)\in {\cal O}^{(N)}$.
Therefore, $\vert  {\cal O}_a^{(N)} \vert=\vert
{\cal O}_0^{(N)}  \vert=\vert{\cal O}^{(N)}\vert=5$ in total.
For example, we have $(1002),(1020),(1200),(3000)$, and $(0111)$ 
in  ${\cal O}_a^{(N)}$.

In fact, every operator at higher $k$ can be generated by
adding 2 to one element of $n_X$ $k$ times or
adding 2 to one element of $\bar n_X$ $k-1$ times.
A number of independent operators at a given $k$, therefore, is obtained by a number of 
selecting $k$ (or $k-1$) numbers from $1,2,3,4$, 
which is $_{k+3}C_3$ ($_{k+2}C_3$). In total we have
\beqa
_{k+3}C_3+_{k+2}C_3 &=& \frac{(k+2)(k+1)(2k+3)}{3!},
\eeqa
for each set. This completes the proof for 
$\vert {\cal O}_a^{(N)}\vert = \vert {\cal O}_0^{(N)}\vert 
=\vert{\cal O}^{(N)}\vert $ at an arbitrary odd number $N$.

\subsection{WTI's at even $N$}
We next consider the case with $N=2k$.
We have $n_A(n_B)=0110$($0011$) and $\bar n_A(\bar n_B)=1001$($1100$) at $k=1$ for ${\cal O}_a^{(N)}$
(${\cal O}_0^{(N)}$), while $n_C=0000$ at $k=0$ and $\bar n_C=1111$ at $k=2$ for ${\cal O}^{(N)}$.
As before, it is easy to count a number of independent operators for each case.
There are $ 2\times( {}_{k+2}C_3)$ operators for ${\cal O}_a^{(N)}$ and ${\cal O}_0^{(N)}$, while there are
${}_{k+3}C_3+{}_{k+1}C_3$ for ${\cal O}^{(N)}$. From this, it is easy to see that
\beqa
\vert {\cal O}^{(N)}\vert - \vert {\cal O}_{a,0}^{(N)}\vert
&=& {}_{k+3}C_3+{}_{k+1}C_3 - 2\times ({}_{k+2}C_3) = (k+1) > 0,
\eeqa
which means that $\vert {\cal O}^{(N)}\vert  > \vert {\cal O}_a^{(N)}\vert =\vert {\cal O}_0^{(N)}\vert$ at $N=2k$. Therefore  $\delta_a  {\cal O}_0^{(N)}=0$ is not equivalent to  $\delta_0  {\cal O}_0^{(N)}=0$.

\subsection{Explicit WTI's at small $N=2k$}
\subsubsection{$k=1$}
In this case two non-singlet WTI is given by
\beqa
(0020)-(0200) &=& 0, \quad  (2000)-(0002) = 0,
\eeqa
while the singlet ones give
\beqa
\delta^0(0011)&=&(0020)-(0002), \\
 \delta^0(1100)&=&(2000)-(0200)= (0002)-(0020)
 =-\delta^0(0011) .
\eeqa
Therefore, one non-trivial $U(1)_A$ rotation can remain.
For simplicity,  we omit the bracket of $\langle (n_1n_2n_3n_4) \rangle $ here.

\subsubsection{$k=2$}
In this case, 8 non-singlet WTI's read
\beqa
(2020) -(2200)- 2(1111) &=& 0, \quad  (4000)-3(2002) = 0, \\
(2200)-(0202)+2(1111) &=&0, \quad 3(0220)-(0400) =0, \\
(2020)-(0022)-2(1111) &=&0, \quad (0040)-3(0220) =0, \\
(0022)-(0202)+2(1111) &=&0, \quad 3(2002)-(0004) =0,  
\eeqa
which can be reduced to
 \beqa
(4000)&=&(0004) = 3(2002), \quad  (0400)=(0040) = 3(0220),\\
(2020)&=& (0202), \quad  (2200)=(0022), \quad 2(1111)=(0202)-(0022).
 \eeqa
 Note that these are 
symmetric under $n_1\leftrightarrow n_4, n_2\leftrightarrow n_3$.

Using the above conditions, we have two independent quantities 
\beqa
(0022) -(2002), \qquad (0022)-(0220),
\eeqa
with which to examine the $U(1)_A$ chiral symmetry.

\subsubsection{$k=3$}
We have 20 WTI's from the non-singlet chiral symmetry: 
\beqa
(4020)-(4200)-4(3111) &=& 0,\quad (6000)-5(4002)=0, \\
(2400)-(0402)+4(1311) &=&0 , \quad (0600)-5(0420)=0, \\
3(2220)-(2400)-2(1311) &=&0,\quad  (4200)-3(2202)+2(3111) = 0, \\
(2040)-3(2220)-2(1131) &=& 0, \quad (4020)-3(2022)-2(3111) =0  
\eeqa 
plus equations derived from the above by 
$n_1\leftrightarrow n_4, n_2\leftrightarrow n_3$, 
and
\beqa
(2022)-(2202) -2(1113)+2(3111) &=& 0, \quad
3(4002)-3(2004) = 0, \\
(2220)-(0222) +2(1131)-2(1311) &=& 0, \quad
3(0240)-3(0420) = 0.
\eeqa
The above conditions are summarized as
\beqa
(4002)&=&(2004), \ (0240)=(0420), \ (4020)=(0204), \nn\\
(4200)&=&(0024), \ (2400)=(0042), \ (0402)=(2040), \nn\\
(6000)&=&(0006) = 5 (2004), \quad
(0600)=(0060)=5(0420), \nn \\
6(2220)&=&6(0222)=(0042)+(2040), \quad
6(2202)=6(2022)=(0204)+(0024), \nn \\
4(3111)&=&4(1113)=(0204)-(0024),\quad
4(1311)=4(1131) =(2040)-(0042). \nn \\
\eeqa
In this case, there remain 4 non-trivial 
chiral $U(1)_A$ rotations as
\beqa
(0024)-(2004), \quad
(2040)-(0204), \quad
 (0420)-(0042), \quad (0042)-(0024).
\eeqa

%-------------------------------------------------------------%

\end{document}